\RequirePackage[2020-02-02]{latexrelease}
\documentclass[aps,prb,twocolumn,floatfix,letterpaper,longbibliography]{revtex4-2}
\usepackage{graphicx}
\usepackage{bm}
\usepackage{amsmath,amsfonts}
\usepackage{amsbsy}
\usepackage[colorlinks=true,linkcolor=blue,urlcolor=blue,citecolor=blue,breaklinks=true]{hyperref}
\usepackage[utf8]{inputenc}
\newcommand{\no}{\!:\!}
\newcommand{\triplecolon}{\raisebox{-0.075ex}{\scalebox{0.7}{\bm\vdots}}}

\def\TITLE{Elementary excitations of a system of one-dimensional
  chiral fermions with short-range interactions}

\begin{document}

\title{\TITLE}

\author{K. A. Matveev}

\affiliation{Materials Science Division, Argonne National Laboratory,
  Argonne, Illinois 60439, USA}

\date{February 3, 2023}

\begin{abstract}

  We study general features of the excitation spectrum of a system of
  one-dimensional chiral spinless fermions with short-range
  interactions.  We show that the nature of the elementary excitations
  of such a system depends strongly on the nonlinearity of the
  underlying dispersion of the fermions.  In the case of quadratic
  nonlinearity, the low-momentum excitations are essentially fermionic
  quasiparticles and quasiholes, whereas the high-momentum ones are
  classical harmonic waves and solitons.  In the case of cubic
  nonlinearity, the nature of the elementary excitations does not
  depend on momentum and is determined by the strength of the
  interactions.  At a certain critical value of the interaction
  strength the excitation spectrum changes qualitatively, pointing to
  a dynamic phase transition in the system.

\end{abstract}
\maketitle

\section{Introduction}
\label{sec:introduction}

The effect of interactions on the low-energy properties of systems of
fermions is strongly enhanced in one dimension.  Indeed, in three
dimensions such a system can be described by the Fermi liquid theory
\cite{lifshitz_statistical_1980}, in which the elementary excitations
are similar to those of the Fermi gas.  On the other hand, the
low-energy properties of systems of interacting one-dimensional
fermions are commonly described in the framework of the Luttinger
liquid theory \cite{haldane_luttinger_1981, giamarchi_quantum_2004},
in which excitations have bosonic statistics.  The best known
signature of the Luttinger liquid behavior is the power-law
renormalization of the tunneling density of states
\cite{kane_transmission_1992, furusaki_single-barrier_1993}.

The issue of the nature of the elementary excitations of the Luttinger
liquid is rather subtle.  At low energies many properties of the
system, including the tunneling density of states, are described by
the Luttinger model, in which the dispersion of the fermions is
approximated by two linear chiral branches
\cite{luttinger_exactly_1963}.  In the absence of interactions, the
many-body spectrum of this model is highly degenerate.  The
Hamiltonian of the system can be presented as that of the system of
the original fermions or as that of noninteracting bosons.
Importantly, the degeneracy is lifted when nontrivial interactions of
the fermions are included in the Luttinger model.  Typical
interactions destroy the picture of free fermionic quasiparticles,
whereas the bosons remain noninteracting, albeit with a nonlinear
dispersion.  Alternatively, the degeneracy of the spectrum can be
lifted by including nonlinear corrections to the dispersion of the
fermions.  This perturbation preserves the picture of free fermionic
quasiparticles, but generates interactions of bosonic excitations.

In a typical physical realization of one-dimensional fermions, the
dispersion is nonlinear and interactions are not negligible.  The
nonlinearity is usually quadratic, $\delta\epsilon_p^{(f)}=p^2/2m$,
where $m$ is the effective mass of the particles at the Fermi level.
On the other hand, for the short-range interactions, the nonlinearity
of the bosonic spectrum is cubic, $\delta\epsilon_p^{(b)}=\zeta p^3$.
At low energies, corresponding to $p\ll p^*\sim(m\zeta)^{-1}$, the
nonlinearity of the fermion dispersion is the dominant perturbation to
the Luttinger model at low energies, and the elementary excitations
are fermions \cite{rozhkov_fermionic_2005}.

The elementary excitations of a system of one-dimensional fermions can
be studied in detail \cite{pustilnik_solitons_2015,
  pustilnik_fate_2015} in the case of strong repulsive interactions,
where the crossover momentum $p^*$ is small compared to the Fermi
momentum $p_F$.  At low energies the Hamiltonian of the system splits
into two chiral Hamiltonians, which can be reduced
\cite{pustilnik_fate_2015} to the quantum Korteweg-de Vries (KdV)
model \cite{sasaki_field_1987, pogrebkov_boson-fermion_2003}.  At
$p\ll p^*$ the system has two branches of elementary excitations near
each Fermi point, with dispersions showing quadratic nonlinearities,
$\delta\varepsilon_p=\pm p^2/2m^*$.  They correspond to the fermionic
quasiparticle and quasihole expected from
Ref.~\cite{rozhkov_fermionic_2005}.  The dispersion undergoes a
crossover at $p\sim p^*$.  At $p\gg p^*$ the quantum KdV Hamiltonian
approaches the classical limit.  In this regime one of the excitation
branches becomes a boson and has a nonlinear dispersion with
$\delta\varepsilon_p\propto p^3$, as expected from the Luttinger model
with short-range interactions.  The second branch corresponds to the
classical KdV soliton; the nonlinearity of its dispersion is
$\delta\varepsilon_p\propto p^{5/3}$.

The goal of this paper is to study the spectrum of the elementary
excitations of a system of interacting chiral spinless fermions, such
as those at the edge of the integer Quantum Hall system
\cite{halperin_quantized_1982} with occupation fraction $\nu=1$.  We
will limit ourselves to the case of short-range interaction and assume
that the dispersion of the fermions has either quadratic or cubic
nonlinearity.  The case of quadratic nonlinearity is relatively
straightforward, because upon bosonization
\cite{haldane_luttinger_1981} the Hamiltonian of the system again
reduces to that of the quantum KdV model \cite{sasaki_field_1987,
  pogrebkov_boson-fermion_2003}.  The case of cubic nonlinearity of
the fermion dispersion, $\delta\epsilon_p^{(f)}=\gamma p^3$, is
qualitatively different.  Most importantly, because the dispersion of
the bosonic excitations of the Luttinger liquid also has cubic
nonlinearity, $\delta\epsilon_p^{(b)}=\zeta p^3$, the relative
significance of the dispersion curvature and interactions does not
depend on momentum.  The crossover between the regimes of fermionic
and bosonic excitations is controlled by the effective interaction
strength $\zeta/\gamma$.  Upon bosonization, the Hamiltonian of the
system reduces to that of the quantum modified KdV (mKdV) model
\cite{pogrebkov_boson-fermion_2003}.  Although the latter is known to
be integrable \cite{sasaki_field_1987}, no exact results for the
dispersion of the elementary excitations are available.  We study the
excitation spectrum of this model numerically and identify the limits
where the elementary excitations are either fermionic quasiparticles
and quasiholes, or semiclassical phonons and solitons.

The paper is organized as follows.  In Sec.~\ref{sec:quadratic} we
study the case of chiral fermions with quadratic dispersion.  In
particular, we demonstrate that the boundaries of the many-body
spectrum coincide with the two branches of the elementary excitations
of the system.  In Sec.~\ref{sec:cubic} we study the case of cubic
dispersion.  We compute the boundaries of its many-body spectrum
numerically and identify the regions where they correspond to the
elementary excitations of the system.  We summarize and discuss our
results in Sec.~\ref{sec:discussion}.

\section{Chiral fermions with quadratic dispersion}
\label{sec:quadratic}

We consider systems of spinless chiral one-dimensional fermions with
two-body interactions.  In general, the Hamiltonian of such a system
has the form
\begin{equation}
  \label{eq:H_general}
  \hat H = \sum_p\epsilon_p\triplecolon c_p^\dagger c_p^{}\triplecolon
  +\frac{1}{L}
  \sum_{\substack{p_1 p_2\\ q>0}}
  V(q)c_{p_1+q}^\dagger c_{p_1}^{} c_{p_2}^\dagger
      c_{p_2+q}^{}.
\end{equation}
Here the operator $c_p$ annihilates a fermion in a state with momentum
$p$ and energy $\epsilon_p$, and $L$ is the system size.  We assume
periodic boundary conditions, and thus $p$ and $q$ are integer
multiples of $2\pi\hbar/L$.  The notation
$\triplecolon\ldots\triplecolon$ indicates normal ordering of the
fermion operators with respect to the ground state $|0\rangle$, in
which all the fermionic states with $p\leq0$ are filled and those with
$p>0$ are empty.

We are interested in the case of short-range interactions, for which
the Fourier transform of the interaction potential $V(q)$ is well
defined at $q=0$ along with its second derivative \cite{twofootnote}.
At sufficiently small $q$ we then approximate
\begin{equation}
  \label{eq:Vshort}
  V(q)=V(0)-\eta q^2,
\end{equation}
where $\eta=-V''(0)/2$.  For typical repulsive interactions $V(0)$ and
$\eta$ are positive.

A single-channel system of chiral fermions has only one Fermi point
$p_F$.  As discussed above, even in the vicinity of the Fermi level it
is important to account for the nonlinearity of the dispersion
$\epsilon_p$.  In this Section, we consider the generic case, in which
the nonlinearity of the dispersion near $p_F$ is quadratic,
\begin{equation}
  \label{eq:quadratic_nonlinearity}
  \epsilon_p=v_F(p-p_F)+\frac{(p-p_F)^2}{2m}.
\end{equation}
Here $v_F$ and $m$ are the Fermi velocity and the effective mass.  To
simplify the treatment of the finite-size effects we define
$p_F=\pi\hbar/L$, which is equidistant from the highest occupied
single particle state $p=0$ and the lowest unoccupied state
$p=2\pi\hbar/L$ in the many-body ground state $|0\rangle$.

To make further progress we bosonize the Hamiltonian given by
Eqs.~(\ref{eq:H_general})--(\ref{eq:quadratic_nonlinearity}) using the
standard expression \cite{haldane_luttinger_1981} for the fermion
annihilation operator at point $x$,
\begin{equation}
  \label{eq:bosonization}
  \Psi(x)=\frac{1}{\sqrt L}
  U \no e^{i\phi(x)}\no.
\end{equation}
The field $\phi(x)$ is expressed in terms of the bosonic
annihilation operators $b_l$ as
\begin{equation}
  \label{eq:phi}
  \phi(x)=2\pi N\frac{x}{L}
  -\sum_{l=1}^\infty
    \frac{i}{\sqrt l}\left(e^{i2\pi l x/L}b_l
    -e^{-i2\pi l x/L}b_l^\dagger\right).
\end{equation}
Here $N$ is the operator of the number of particles measured from that
in the ground state $|0\rangle$; the operator $U$ lowers the number of
particles by 1, i.e., $[U,N]=U$.  The colons in
Eq.~(\ref{eq:bosonization}) denote the normal ordering of the bosonic
operators $b_l^{}$ and $b_l^\dagger$.

Expressing the fermion operators $c_p$ in the Hamiltonian
(\ref{eq:H_general}) in terms of the bosonic field $\phi(x)$ with the
help of Eq.~(\ref{eq:bosonization}) and substituting
Eqs.~(\ref{eq:Vshort}) and (\ref{eq:quadratic_nonlinearity}), we
obtain
\begin{equation}
  \label{eq:H_quadratic}
  \hat H = v(\hat P-p_FN)
  -\frac{\pi^2\hbar^2N}{6mL^2}+\hat H_{\rm KdV}.
\end{equation}
Here $v=v_F+V(0)/2\pi\hbar$, and $\hat P$ is the operator of the total
momentum of the system,
\begin{eqnarray}
  \label{eq:P}
  \hat P &=&
  p_FN+\frac{\hbar}{4\pi}\int\limits_{-L/2}^{L/2}\no(\partial_x\phi)^2\no dx
\nonumber\\
         &=&\frac{\pi\hbar N(N+1)}{L}
             +\sum_{l=1}^\infty\frac{2\pi\hbar l}{L}b_l^\dagger b_l^{},
\end{eqnarray}
measured from that of the ground state $|0\rangle$.
The last term in Eq.~(\ref{eq:H_quadratic}) is given by
\begin{equation}
  \label{eq:H_kdv}
  \hat H_{\rm KdV}=
  \frac{\hbar^2}{12\pi m}
  \int\limits_{-L/2}^{L/2}\!\!\!
  \no[(\partial_x\phi)^3-a^*(\partial_x^2\phi)^2]\no dx,
  \quad
  a^*=\frac{3m\eta}{2\pi}.
\end{equation}
Equations (\ref{eq:H_quadratic})--(\ref{eq:H_kdv}) fully account for
the effects of changing particle number in the system.  In the
following, we limit ourselves to the sector of the Hilbert space
corresponding to a fixed number of particles.  In this case, without
loss of generality one can set $N=0$, resulting in
\begin{equation}
  \label{eq:H_quadratic_simple}
  \hat H = v \hat P+\hat H_{\rm KdV}.
\end{equation}
Since the Hamiltonian (\ref{eq:H_general}) conserves momentum, we
conclude that $\hat H$ and $\hat H_{\rm KdV}$ have common eigenstates
and that the corresponding energies of any state with momentum $p$
are related by
\begin{equation}
  \label{eq:energy_relation}
  \varepsilon(p)=vp+\varepsilon_{\rm KdV}^{}(p),
\end{equation}
where $\varepsilon$ and $\varepsilon_{\rm KdV}$ are the eigenvalues of
the Hamiltonians $\hat H$ and $\hat H_{\rm KdV}$, respectively.

\begin{figure}[t]
\includegraphics[width=.4\textwidth]{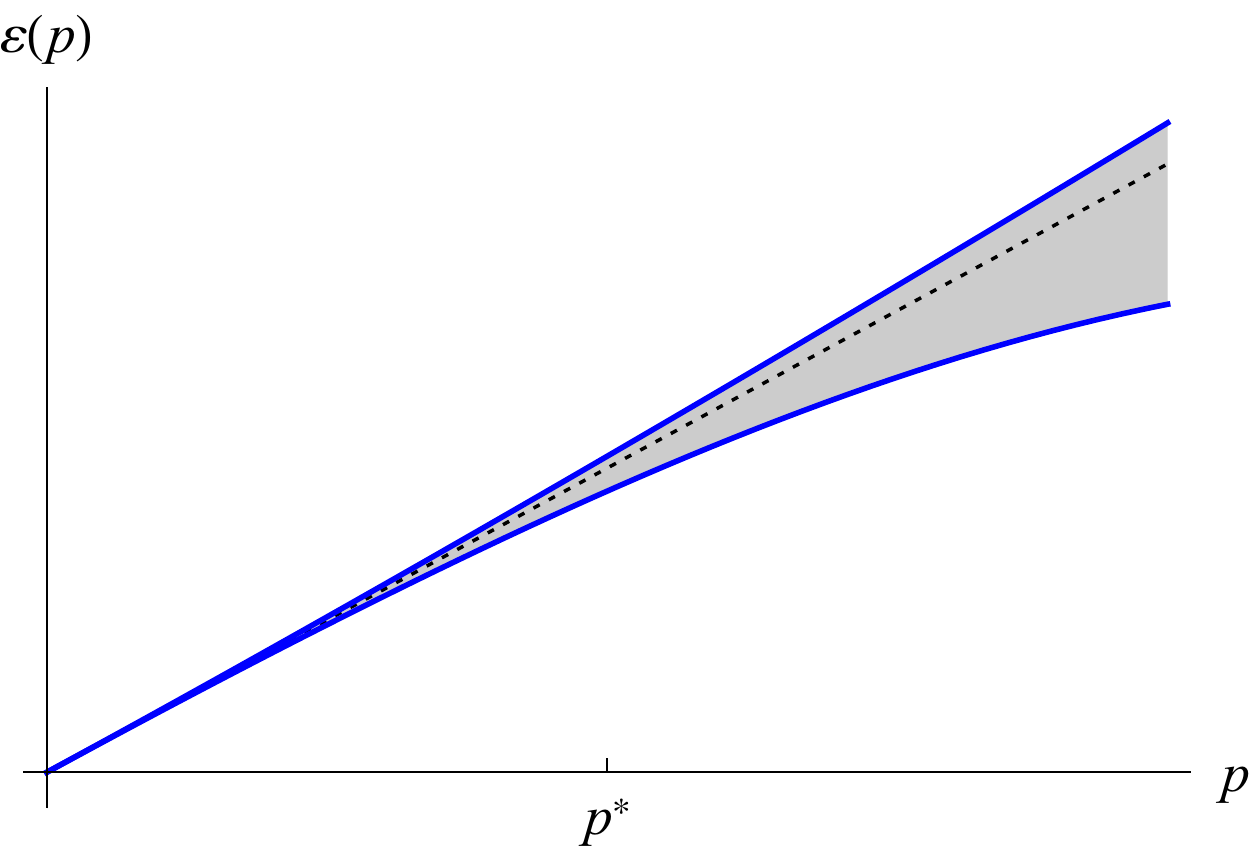}
\caption{Dispersions of the two branches of the elementary excitations
  of the system of interacting chiral fermions with quadratic
  dispersion. The two solid lines are obtained by combining
  Eqs.~(\ref{eq:energy_relation}) and (\ref{eq:KdV_dispersions}), with
  the (arbitrarily chosen) velocity value $v=10p^*/m$.  The dotted line
  shows the linear part of the dispersion relation,
  $\varepsilon(p)=vp$.  All the eigenstates of the system are confined
  to the shaded region between the solid lines.}
\label{fig:dispersions_plot}
\end{figure}

In combination with the commutation relation
\begin{equation}
  \label{eq:commutator}
  [\phi(x),\partial_y\phi(y)]=-2\pi i\delta(x-y),
\end{equation}
which follows from Eq.~(\ref{eq:phi}), the Hamiltonian
(\ref{eq:H_kdv}) defines the quantum KdV problem
\cite{sasaki_field_1987}.  It can also be obtained
\cite{pustilnik_fate_2015} by applying a certain limiting procedure to
the bosonized Hamiltonian of either the Lieb-Liniger model
\cite{lieb_exact_1963}, or the hyperbolic Calogero-Sutherland model
\cite{sutherland_brief_1978, sutherland_beautiful_2004}.  The latter
two models have Bethe ansatz solutions, and their elementary
excitations are well understood.  Using those results, the exact
dispersions of the elementary excitations of the quantum KdV model at
$L\to\infty$ have been obtained in Ref.~\cite{pustilnik_fate_2015},
\begin{equation}
  \label{eq:KdV_dispersions}
  \varepsilon_{\rm KdV}(p)=\frac{{p^*}^2}{2m}e_\pm\left(\frac{p}{p^*}\right).
\end{equation}
Here $p^*=3\hbar/2a^*$, which in our notations takes the form
\begin{equation}
  \label{eq:p^*}
  p^*=\frac{\pi\hbar}{m\eta}.
\end{equation}
The dimensionless functions $e_+(s)$ and $e_-(s)$ correspond to the
two branches of the elementary excitations.  Their exact analytic
expressions (in quadratures) can be found in
Ref.~\cite{pustilnik_fate_2015}; here we quote their limiting
behaviors at large and small $s$.  For the upper branch we have
\begin{equation}
  \label{eq:e_+_limits}
  e_+(s)
  =\left\{
      \begin{array}[c]{ll}
        s^2,& s\ll1
        \\[1ex]
        \frac{3}{5}\left(\frac{2\pi}{3}\right)^{2/3}s^{5/3},& s\gg1,
      \end{array}
     \right.
\end{equation}
whereas for the lower one,
\begin{equation}
  \label{eq:e_-_limits}
  e_-(s)
  =\left\{
      \begin{array}[c]{ll}
        -s^2,& s\ll1
        \\[1ex]
        -s^3,& s\gg1.
      \end{array}
      \right.
\end{equation}
The dispersions of the elementary excitations are illustrated in
Fig.~\ref{fig:dispersions_plot}.

The above asymptotic expressions for the energies of the elementary
excitations allow for simple physical interpretations.  The scaling
dimensions of the first and second terms in the Hamiltonian density in
Eq.~(\ref{eq:H_kdv}) are $3$ and $4$, respectively.  Thus at small
momentum the second term, which accounts for the interactions of
fermions, can be neglected.  The remaining term accounts for the
quadratic nonlinearity of the dispersion of the fermions.  Indeed, at
$p\ll p^*$, Eqs.~(\ref{eq:KdV_dispersions}), (\ref{eq:e_+_limits}),
and (\ref{eq:e_-_limits}) yield quasiparticle dispersions
$\varepsilon_{\rm KdV}=\pm p^2/2m$, which correspond to the fermionic
quasiparticle and quasihole excitations.

At large momentum $p\gg p^*$, the system is in the interaction
dominated regime.  Substituting Eq.~(\ref{eq:phi}) into the second
term in Eq.~(\ref{eq:H_kdv}), we bring the interaction Hamiltonian to
the form
\begin{equation}
  \label{eq:H_int}
  H_{\rm int}=-\frac{\eta}{2\pi\hbar}\sum_{l=1}^\infty
  p_l^3b_l^\dagger b_l^{},
  \quad
  p_l=\frac{2\pi\hbar}{L}l.
\end{equation}
Thus at large momentum $p\gg p^*$, the nonlinear correction to the
energy of the bosonic excitations in the Luttinger liquid
approximation scales as $-p^3$, in agreement with the second line of
Eq.~(\ref{eq:e_-_limits}).  To obtain the physical interpretation of
the other mode, one should restore the small first term in
Eq.~(\ref{eq:H_kdv}) and derive the equation of motion for the
particle density operator $\partial_x\phi/2\pi$.  At $p\gg p^*$ and
$L\to\infty$, it becomes the classical KdV equation
\cite{pustilnik_fate_2015}.  The latter has two types of solutions.
First, there are solutions representing harmonic waves of
infinitesimal amplitude, which have cubic dependence of frequency on
the wavevector.  They are equivalent to the bosonic excitations in
Eq.~(\ref{eq:H_int}).  Second, there are soliton solutions of the KdV
equation, for which the energy scales as $p^{5/3}$
\cite{pustilnik_fate_2015}.  They correspond to the second line of
Eq.~(\ref{eq:e_+_limits}).

Let us now discuss the full many-body energy spectrum of the
Hamiltonian (\ref{eq:H_kdv}).  In the absence of interactions,
$\eta=0$, each of the two simplest many-body states with momentum $p$
contains only a single elementary excitation, quasiparticle or
quasihole, and the corresponding energies are $p^2/2m$ and $-p^2/2m$,
respectively.  Since any state of the system can be viewed as a
combination of particles and holes, and all excitations of a chiral
system have positive momenta, the quasiparticle energies $\pm p^2/2m$
are the highest and lowest energies possible at the total momentum
$p$.  Because the Bethe ansatz eigenstates at $\eta\neq0$ can also be
classified by occupation numbers of quasiparticles and quasiholes, it
is natural to expect \cite{pustilnik_solitons_2015} that the two
branches of elementary excitations (\ref{eq:KdV_dispersions})
represent exact boundaries of the many-body spectrum at any
interaction strength.  This argument applies to many-body states with
only a few quasiparticles and quasiholes, because their interactions
vanish in the limit of infinite system size.  A generic state,
however, has a finite density of quasiparticles and quasiholes, and
the above simple argument does not apply.  Nevertheless, we conjecture
that Eq.~(\ref{eq:KdV_dispersions}) yields the exact boundaries of the
excitation spectrum at any $\eta$.  For the system of interacting
chiral fermions defined by
Eqs.~(\ref{eq:H_general})--(\ref{eq:quadratic_nonlinearity}), this
means that the full energy spectrum is confined to the shaded region
in Fig.~\ref{fig:dispersions_plot}.

\begin{figure}[t]
\includegraphics[width=.45\textwidth]{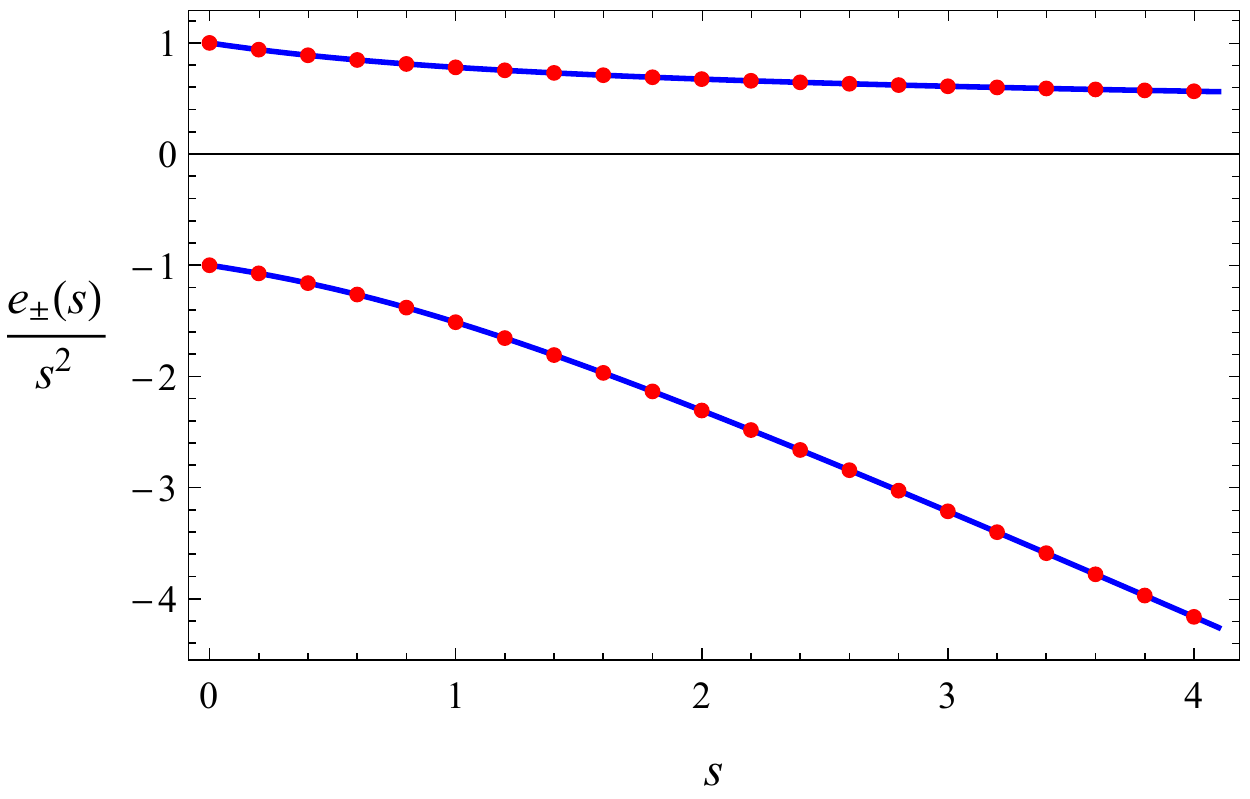}
\caption{Comparison of the quasiparticle energies
  (\ref{eq:KdV_dispersions}) with the spectral boundaries of the
  Hamiltonian (\ref{eq:H_kdv}).  The horizontal axis $s$ is the
  momentum in units of $p^*$ given by Eq.~(\ref{eq:p^*}).  The solid
  lines are the energies (\ref{eq:KdV_dispersions}) divided by
  $p^2/2m$.  The dots represent the numerically computed boundaries of
  the energy spectrum of the Hamiltonian (\ref{eq:H_kdv}) extrapolated
  to infinite system size.}
\label{fig:KdV-numerics}
\end{figure}

We now verify the above conjecture by numerical diagonalization of the
quantum KdV Hamiltonian (\ref{eq:H_kdv}).  In
Fig.~\ref{fig:KdV-numerics} we plot the upper and lower boundaries of
the spectrum normalized by $p^2/2m$.  The conjectured values of the
boundaries are given by Eq.~(\ref{eq:KdV_dispersions}), which upon
normalization yields $e_\pm(s)/s^2$, where $s=p/p^*$.  We used the
analytic expressions for $e_\pm(s)$ obtained in
Ref.~\cite{pustilnik_fate_2015} and plotted $e_\pm(s)/s^2$ as solid
lines in Fig.~\ref{fig:KdV-numerics}.  The horizontal axis represents
$s=p/p^*=pm\eta/\pi\hbar$, see Eq.~(\ref{eq:p^*}).  In the numerical
calculation, we fix the total momentum $p$ and find the highest and
lowest eigenvalues of the Hamiltonian (\ref{eq:H_kdv}) for different
values of the interaction strength $\eta$.  The eigenvalues are then
divided by $p^2/2m$ and the results are extrapolated numerically to
the limit of infinite system size \cite{onefootnote}.  The resulting
spectral boundaries are shown by dots in Fig.~\ref{fig:KdV-numerics}.
They are in excellent agreement with the conjectured values.

To summarize the results of this Section, we have studied the
elementary excitations of the system of interacting chiral spinless
fermions described by
Eqs.~(\ref{eq:H_general})--(\ref{eq:quadratic_nonlinearity}).  Upon
bosonization, the Hamiltonian reduces to that of the quantum KdV
model, see Eq.~(\ref{eq:H_kdv}).  At low momentum, the excitations are
the fermionic quasiparticle and quasihole.  At high momentum they
become the KdV soliton and the harmonic wave.  The energies are given
by Eqs.~(\ref{eq:energy_relation}) and (\ref{eq:KdV_dispersions}),
where the exact expressions for the functions $e_\pm(s)$ can be found
in Ref.~\cite{pustilnik_fate_2015}.  The two quasiparticle branches
are shown in Fig.~\ref{fig:dispersions_plot}.  They represent the
exact boundaries of the many-body spectrum of the system.

\section{Chiral fermions with cubic dispersion}
\label{sec:cubic}

\subsection{Hamiltonian}
\label{sec:cubic_Hamiltonian}

We now turn to the special regime of the interacting system of
spinless chiral fermions, in which the Fermi momentum $p_F$ coincides
with an inflection point of the dispersion $\epsilon_p$.  In this case
the effective mass $m$ is infinite, and $\epsilon_p$ has a cubic
nonlinearity near the Fermi point,
\begin{equation}
  \label{eq:cubic_nonlinearity}
  \epsilon_p=v_F(p-p_F)+\gamma(p-p_F)^3.
\end{equation}
In the following, we assume $\gamma>0$.  The results for negative
$\gamma$ can be obtained by trivial symmetry transformations.

Next, we apply the bosonization transformation (\ref{eq:bosonization})
to bring the Hamiltonian given by Eqs.~(\ref{eq:H_general}),
(\ref{eq:Vshort}), and (\ref{eq:cubic_nonlinearity}) to the form
\begin{equation}
  \label{eq:H_cubic}
  \hat H = \left(v-\frac{\pi^2\hbar^2\gamma}{L^2}\right)(\hat P-p_FN)
  +\hat H_{\rm mKdV}.
\end{equation}
Here again $v=v_F+V(0)/2\pi\hbar$, the operator of the total momentum
is defined by Eq.~(\ref{eq:P}), and we again chose the Fermi momentum
$p_F=\pi\hbar/L$.  The last term is given by
\begin{equation}
  \label{eq:H_mkdv}
  H_{\rm mKdV}=
  \frac{\hbar^3\gamma}{8\pi}
    \int\limits_{-L/2}^{L/2}\!\!\!
  \no[(\partial_x\phi)^4+(1-2\chi)(\partial_x^2\phi)^2]\no dx,
\end{equation}
where $\chi=\eta/2\pi\hbar\gamma$.  In combination with the
commutation relation (\ref{eq:commutator}), the Hamiltonian
(\ref{eq:H_mkdv}) defines the quantum mKdV problem
\cite{sasaki_field_1987}.  In the limit of infinite system size, the
relation between the quantum mKdV Hamiltonian and that of chiral
fermions was discussed in Ref.~\cite{pogrebkov_boson-fermion_2003}.

Unlike the case of quadratic nonlinearity
(\ref{eq:quadratic_nonlinearity}), the Hamiltonian given by
Eqs.~(\ref{eq:H_general}), (\ref{eq:Vshort}), and
(\ref{eq:cubic_nonlinearity}) possesses particle-hole symmetry, i.e.,
it retains its form under the transformation
\begin{equation}
  \label{eq:particle-hole_fermion}
  c_{p_F+q}^{}\to c_{p_F-q}^\dagger,
  \quad
  \Psi(x)\to\Psi^\dagger(x)e^{i2\pi x/L}.
\end{equation}
In the bosonized form, this property is expressed as 
\begin{equation}
  \label{eq:particle-hole_boson}
  U\to U^\dagger,
  \quad
  \phi(x)\to-\phi(x),
  \quad
  b_l\to -b_{l},
  \quad
  N\to-N,
\end{equation}
see Eqs.~(\ref{eq:bosonization}) and (\ref{eq:phi}).

Since the Hamiltonian $\hat H$ conserves the total momentum $\hat P$,
it follows from Eq.~(\ref{eq:H_cubic}) that it has the same
eigenstates at $H_{\rm mKdV}$.  In the sector of the Hilbert space
with the number of particles equal to that in the ground state
$|0\rangle$ one should set $N=0$.  In this case, the eigenvalues of
the two Hamiltonians are related by
\begin{equation}
  \label{eq:energy_relation_mkdv}
  \varepsilon(p)=vp+\varepsilon_{\rm mKdV}^{}(p),
\end{equation}
where we omitted the finite size correction to the velocity $v$ in
Eq.~(\ref{eq:H_cubic}).

\subsection{Boundaries of the energy spectrum}
\label{sec:boundaries}

Our next goal is to study the boundaries of the energy spectrum of the
quantum mKdV Hamiltonian (\ref{eq:H_mkdv}) in the limit of infinite
system size, $L\to\infty$.  We start with the case of free fermions,
corresponding to $\chi=0$.  In this case, the system supports two
types of elementary excitations, the quasiparticles and quasiholes.
The particle-hole symmetry of the Hamiltonian (\ref{eq:H_mkdv})
ensures that the quasiparticles and quasiholes with momentum $p$ have
the same energy $\gamma p^3$.  The latter is the highest energy of any
eigenstate of the Hamiltonian (\ref{eq:H_mkdv}) with momentum $p$ at
$\chi=0$.  The lowest possible energy at $L\to\infty$ is 0.  To obtain
an eigenstate with energy near this lower boundary, the total momentum
$p$ should be divided among many particle-hole pairs, each carrying a
very small fraction of the total momentum.

In addition to the case of vanishing interactions, $\chi=0$, the
Hamiltonian (\ref{eq:H_mkdv}) can be easily treated analytically in
the limits of strong repulsive or attractive interactions,
$\chi\to\pm\infty$.  In this case, the second term in
Eq.~(\ref{eq:H_mkdv}) dominates.  Using Eq.~(\ref{eq:phi}), the latter
can be written in terms of the bosonic operators $b_l$,
\begin{equation}
  \label{eq:H_mkdv_strong}
  H_{\rm mKdV}\simeq
  \gamma\left(\frac{1}{2}-\chi\right)\sum_{l=1}^\infty
  p_l^3b_l^\dagger b_l^{},
  \quad
  p_l=\frac{2\pi\hbar}{L}l,
\end{equation}
cf.~Eq.~(\ref{eq:H_int}).  A system described by the Hamiltonian
(\ref{eq:H_mkdv_strong}) has bosonic elementary excitations with
energies $\gamma(\frac12-\chi)p^3$.  This expression also yields the
lower (upper) boundary of the full energy spectrum at positive
(negative) $\chi-\frac12$.  The other boundary of the spectrum at
$L\to\infty$ is at zero energy.  It corresponds to the state in which
the total momentum $p$ of the system is distributed among an infinite
number of bosons with infinitesimal momentum.

We showed so far that at $\chi=0$ and $\chi\to\pm\infty$, the
eigenstates of the Hamiltonian (\ref{eq:H_mkdv}) with a given momentum
$p$ are confined to regions with the boundaries that scale as $p^3$.
Because both contributions to the Hamiltonian density in
Eq.~(\ref{eq:H_mkdv}) have scaling dimension 4, this observation holds
for any value of $\chi$.  Furthermore, the form of the Hamiltonian
(\ref{eq:H_mkdv}) ensures that for any given $\chi$ all the energies
are proportional to $\gamma$.  Thus the upper and lower boundaries of
the spectrum can be presented in the form
\begin{equation}
  \label{eq:mkdv_boundaries_general}
  \varepsilon_{\rm mKdV}(p)=\alpha_\pm(\chi)\gamma p^3.
\end{equation}
The above results for $\chi=0$ and $\chi\to\pm\infty$ can be
summarized as
\begin{equation}
  \label{eq:alpha_+_limits}
  \alpha_+(\chi)=
  \left\{
    \begin{array}[c]{ll}
      \frac{1}2-\chi,& \chi\to-\infty,
      \\[1ex]
      1,& \chi=0,
      \\[1ex]
      0,& \chi\to+\infty,
      \end{array}
      \right.
\end{equation}
and
\begin{equation}
  \label{eq:alpha_-_limits}
  \alpha_-(\chi)=
  \left\{
    \begin{array}[c]{ll}
      0,& \chi\to-\infty,
      \\[1ex]
      0,& \chi=0,
      \\[1ex]
      \frac{1}2-\chi,& \chi\to+\infty.
      \end{array}
      \right.
\end{equation}

\begin{figure}[t]
\includegraphics[width=.45\textwidth]{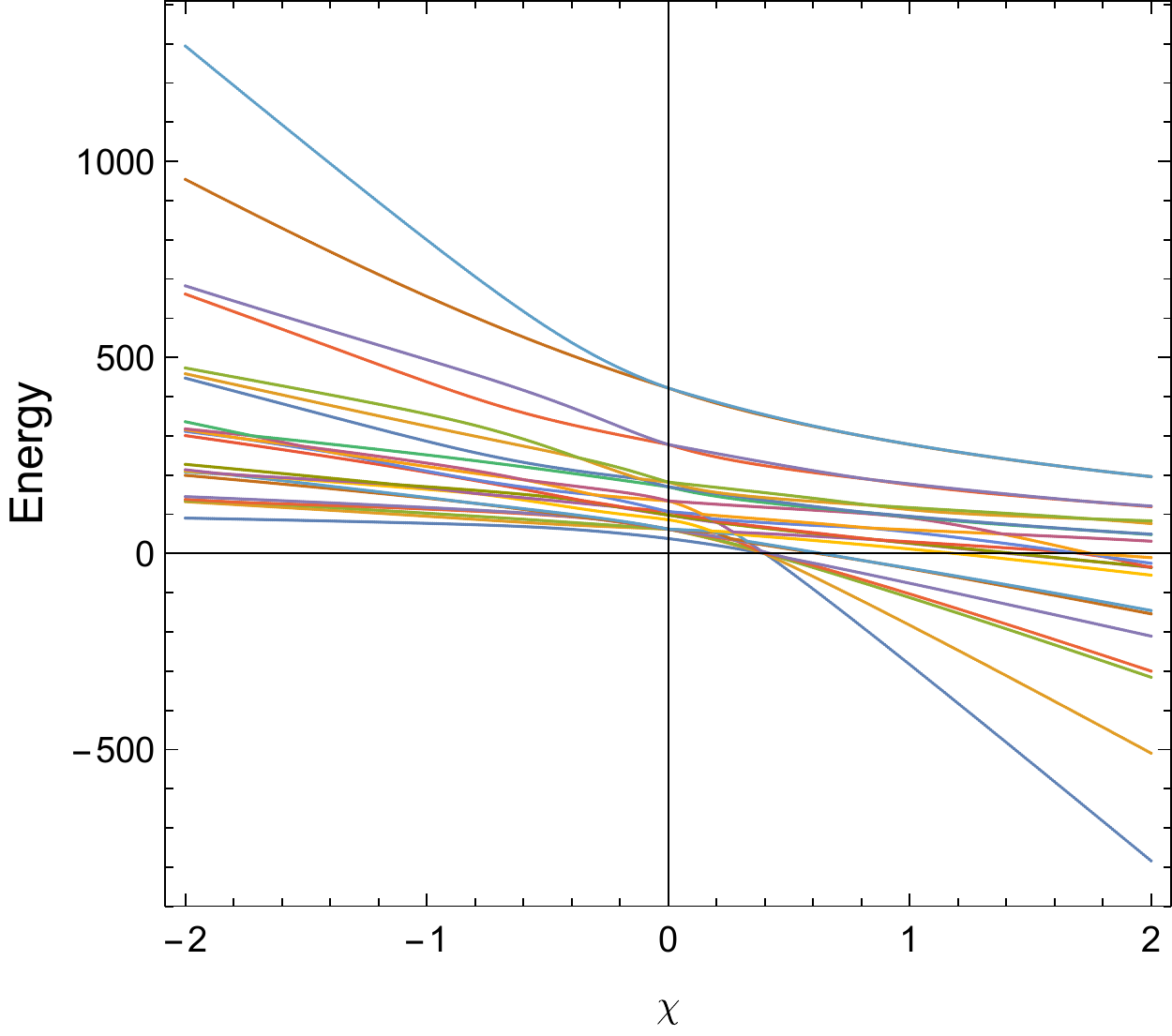}
\caption{The energy spectrum of the Hamiltonian (\ref{eq:H_mkdv}) for
  the total momentum $P=8$ in units of $2\pi\hbar/L$.  The units are
  chosen so that $\gamma=\hbar=1$ and $L=2\pi$.  The solid lines show
  the positions of the 22 energy levels for each value of the
  parameter $\chi$.  (We have subtracted $P/4=2$ from all eigenvalues
  to account for the term $-(\pi^2\hbar^2\gamma/L^2)\hat P$ in
  Eq.~(\ref{eq:H_cubic}) \cite{onefootnote}.)  At $\chi=0$ the two
  highest energy states are degenerate, corresponding to the single
  quasiparticle and quasihole with energies
  $\left(P-\frac12\right)^3+\frac18=422$.}
\label{fig:all-energies}
\end{figure}

\begin{figure}[t]
\includegraphics[width=.45\textwidth]{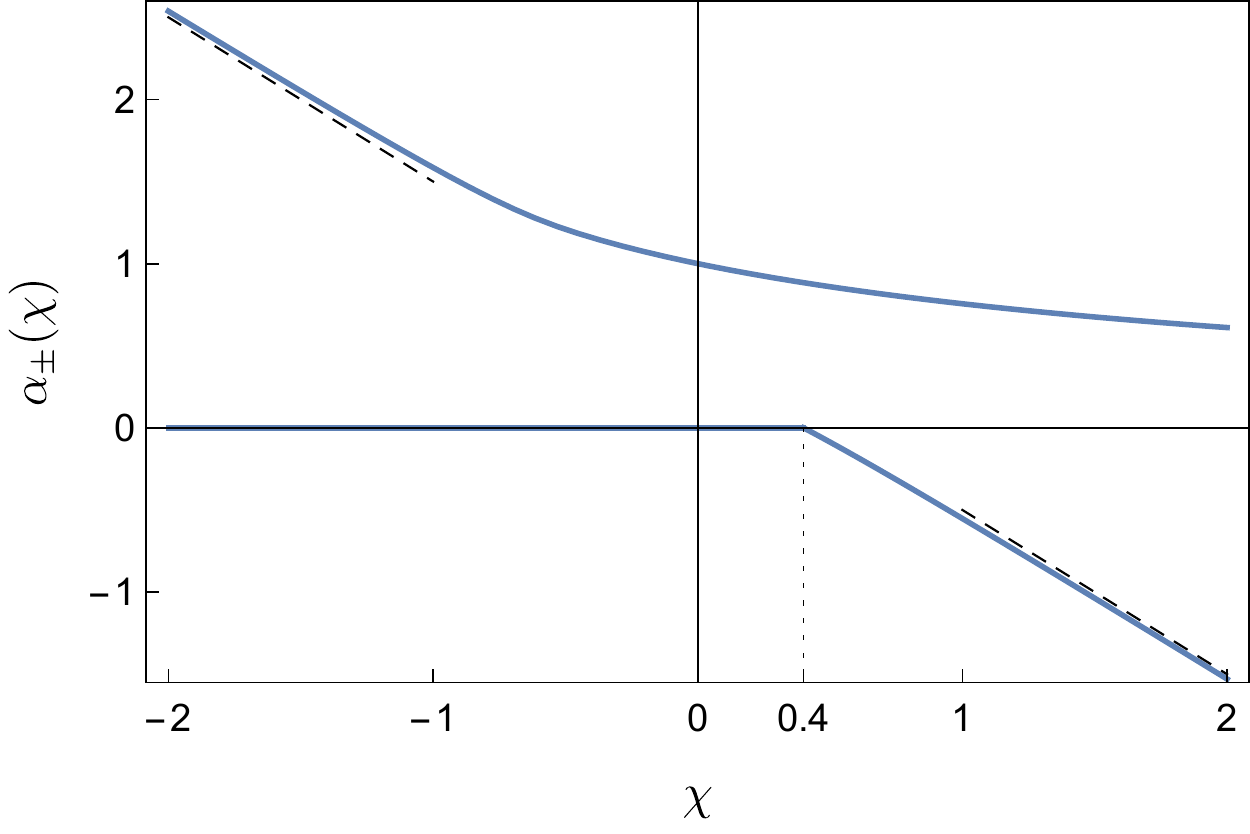}
\caption{The functions $\alpha_+(\chi)$ and $\alpha_-(\chi)$ evaluated
  numerically \cite{onefootnote} are shown by the top and bottom solid
  lines, respectively.  The dashed lines represent the result
  $\frac12-\chi$ for the bosonic excitation branch,
  Eq.~(\ref{eq:H_mkdv_strong}), which controls one of the boundaries
  at large $|\chi|$, see Eqs.~(\ref{eq:alpha_+_limits}) and
  (\ref{eq:alpha_-_limits}).  Note that at $\chi<\chi_c^{}\approx0.4$
  we have $\alpha_-(\chi)=0$.}
\label{fig:alpha-chi}
\end{figure}

The Hamiltonian (\ref{eq:H_mkdv}) can be diagonalized numerically for
moderate values of the total momentum $p$.  The results for the full
energy spectrum in the case of $p=8\times2\pi\hbar/L$ are shown in
Fig.~\ref{fig:all-energies}.  The functions $\alpha_+(\chi)$ and
$\alpha_-(\chi)$ can then be computed by extrapolating the energies of
the highest and lowest levels to the limit $L\to\infty$ at fixed $p$,
see Fig.~\ref{fig:alpha-chi}.

The function $\alpha_-(\chi)$ shows two distinct types of behavior.
When the dimensionless interaction parameter is large and positive
($\chi\to+\infty$), $\alpha_-$ scales as $\frac12-\chi$, in agreement
with Eq.~(\ref{eq:alpha_-_limits}).  As we discussed, the
corresponding lowest energy state of the system at a given value $p$
of the total momentum has a single bosonic excitation with momentum
$p$.  As $\chi$ is lowered, $\alpha_-(\chi)$ increases until it
reaches zero at $\chi=\chi_c^{}\approx 0.4$.  At $\chi<\chi_c^{}$ the
lowest possible energy of the system remains zero.  As we saw earlier,
this is indeed the case for $\chi=0$ and $\chi\to-\infty$, see
Eq.~(\ref{eq:alpha_-_limits}).  The lowest energy state in these cases
corresponded to the total momentum $p$ being distributed among an
infinite number of elementary excitations with infinitesimal momentum.
The numerical results of Fig.~\ref{fig:alpha-chi} imply that this is
the case for all $\chi<\chi_c^{}$.

The behavior of the function $\alpha_+(\chi)$, which describes the
upper boundary of the spectrum, is qualitatively different.  It scales
as $\frac12-\chi$ at large negative $\chi$, which corresponds to the
energy of the system with a single bosonic excitation, see
Eqs.~(\ref{eq:H_mkdv_strong}) and (\ref{eq:alpha_+_limits}).  It
decreases gradually as $\chi$ increases.  At $\chi=0$ it reaches the
value $\alpha(0)=1$, corresponding to a state of free chiral Fermi gas
with a single quasiparticle or quasihole excitation.  It continues to
decrease at positive $\chi$ and gradually approaches zero at
$\chi\to+\infty$, see Eq.~(\ref{eq:alpha_+_limits}).  The nature of
the highest energy state at $\chi\gg1$ is not self-evident, but our
discussion in Sec.~\ref{sec:quadratic} suggests that it may be related
to the classical soliton solutions of the modified KdV equation.

\subsection{Solitons}
\label{sec:solitons}

To study the classical limit of the Hamiltonian (\ref{eq:H_mkdv}), we
use the commutation relation (\ref{eq:commutator}) to write the
equation of motion for the operator $\phi(x,t)$,
\begin{equation}
  \label{eq:phidot}
  \frac{1}{\hbar^2\gamma}\partial_t\phi
  =-(\partial_x\phi)^3
  +\left(\frac12-\chi\right)\partial_x^3\phi.
\end{equation}
In the classical limit, $\phi(x,t)$ is no longer an operator, and thus
we ignored normal ordering.  The conditions under which this
approximation is applicable will be established later.

Next, we introduce $\tau=\hbar^2\gamma t$, differentiate
Eq.~(\ref{eq:phidot}) with respect to $x$, and obtain a partial
differential equation for the function
\begin{equation}
  \label{eq:u}
  u(x,\tau)=\partial_x\phi(x,\tau)
\end{equation}
in the form
\begin{equation}
  \label{eq:mkdv}
  \partial_\tau u + 3u^2\partial_xu
  + \tilde\chi\partial_x^3u=0,
  \quad
  \tilde\chi=\chi-\frac12.
\end{equation}
This is the well-known classical modified KdV equation
\cite{lamb_elements_1980}.  At $\tilde\chi>0$ it has two soliton
solutions
\begin{equation}
  \label{eq:soliton}
  u(x,\tau)=\pm
  \frac{\sqrt{2c}}{\cosh\left(\sqrt{\frac{c}{\tilde\chi}}(x-c\tau)\right)},
\end{equation}
where $c>0$.  No soliton solutions exist for $\tilde\chi<0$.

We are now in a position to obtain the condition of applicability of
the classical approximation used in deriving Eq.~(\ref{eq:phidot}).
Equation (\ref{eq:soliton}) yields the order of magnitude estimate
$u\sim\sqrt{c}$.  Given that the spatial scale of the soliton solution
is $\sqrt{\tilde\chi/c}$, Eq.~(\ref{eq:u}) yields
$\phi \sim u\sqrt{\tilde\chi/c} \sim \sqrt{\tilde\chi}$.  According to
Eq.~(\ref{eq:commutator}), the commutator of $\phi(x)$ and $\phi(y)$
is of order unity.  Classical approximation assumes that it is small
compared to $\phi(x)\phi(y)\sim\tilde\chi$.  Thus the classical
results are applicable at $|\tilde\chi|\gg1$ or, equivalently,
$|\chi|\gg1$.

Integrating the standard bosonization expression for the particle
density $n(x)=\partial_x\phi/2\pi=u/2\pi$ with respect to $x$, we find
the total number of fermions carried by a single soliton
\begin{equation}
  \label{eq:Ns}
  N_s=\pm\sqrt{\frac{\tilde\chi}{2}}.
\end{equation}
Thus, at $\tilde\chi\gg1$, when the classical mKdV theory applies to
the chiral Fermi system with cubic dispersion, the soliton carries a
large number of particles, $|N_s|\gg1$.

Next, we substitute Eq.~(\ref{eq:u}) into Eq.~(\ref{eq:P}) and
(\ref{eq:H_mkdv}) to obtain the expressions for the momentum and
energy of the soliton (\ref{eq:soliton}) in the limit $L\to\infty$.
This yields
\begin{eqnarray*}
  p&=&\frac{\hbar}{4\pi}\int_{-\infty}^{\infty}u^2 dx
       =\frac{\hbar}{\pi}\sqrt{c\tilde\chi},
  \\
  E&=&
       \frac{\hbar^3\gamma}{8\pi}
       \int_{-\infty}^{\infty}\!\!
       \Big(u^4+(1-2\chi){u'}^2\Big)dx
       =\frac{\hbar^3\gamma}{3\pi}\sqrt{c^3\tilde\chi}.
\end{eqnarray*}
These expressions give the dispersion relation for the soliton in the
form
\begin{equation}
  \label{eq:soliton_dispersion}
  E(p)=\frac{\pi^2}{3\tilde\chi}\gamma p^3.
\end{equation}
As expected, the energy scales with the third power of momentum.

\begin{figure}[t]
\includegraphics[width=.45\textwidth]{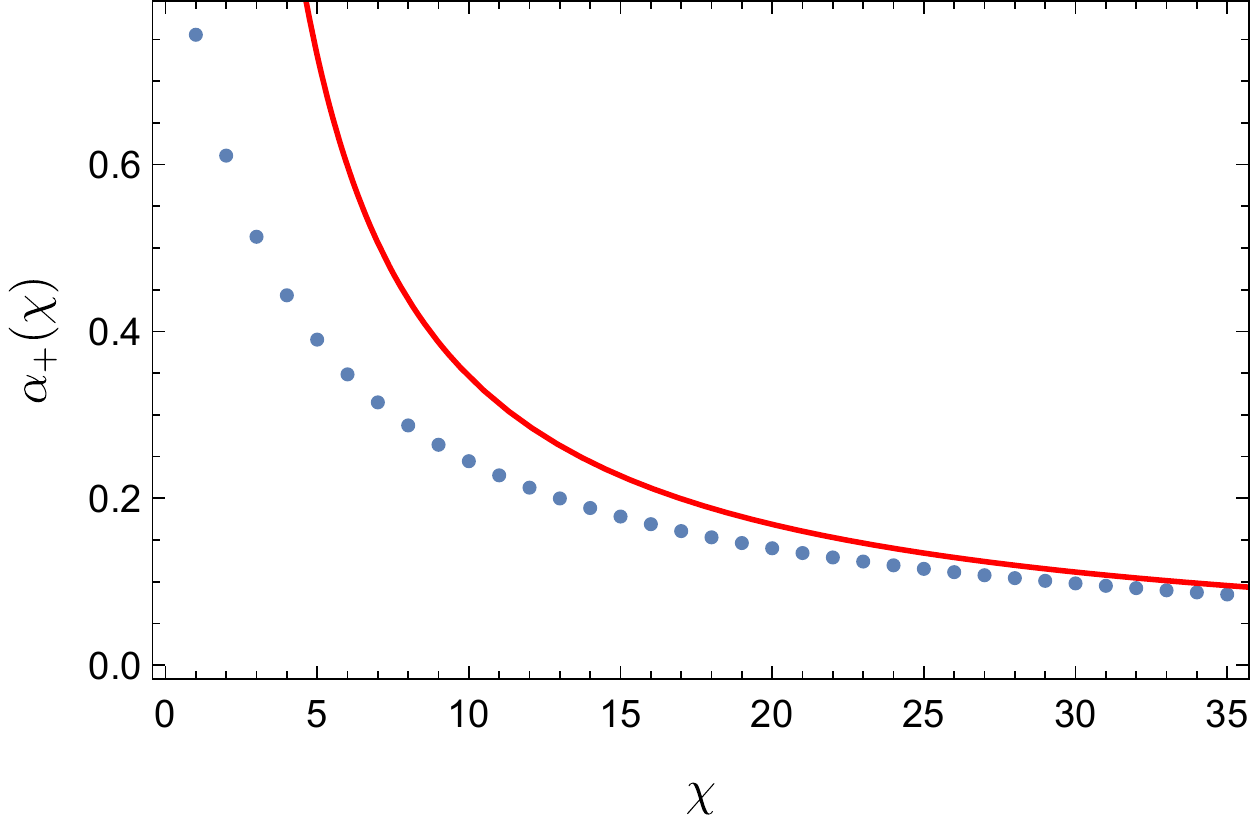}
\caption{Comparison of the numerically computed $\alpha_+(\chi)$ for
  $\chi=1,2,\ldots,35$ (dots) with the asymptotic behavior
  (\ref{eq:alpha_+_soliton}) shown by the solid line.}
\label{fig:soliton-comparison}
\end{figure}

Similarly to the case of a system of chiral fermions with quadratic
dispersion discussed in Sec.~\ref{sec:quadratic}, soliton behaves as
an elementary excitation of the system.  Although states involving
many solitons are possible, the convex shape of the dispersion
relation (\ref{eq:soliton_dispersion}) suggests that the state with
one soliton has the largest energy at a given momentum.  This implies
the following behavior of $\alpha_+(\chi)$ at large $\chi$,
\begin{equation}
  \label{eq:alpha_+_soliton}
  \alpha_+(\chi)\simeq\frac{2\pi^2}{3(2\chi-1)},
  \quad
  \chi\gg1,
\end{equation}
where we used Eqs.~(\ref{eq:mkdv_boundaries_general}),
(\ref{eq:soliton_dispersion}), and the definition of $\tilde\chi$ from
Eq.~(\ref{eq:mkdv}).  The asymptotic behavior
(\ref{eq:alpha_+_soliton}) is consistent with our earlier expectation
of the limiting value at $\chi\to+\infty$ in
Eq.~(\ref{eq:alpha_+_limits}).  It can also be compared with the
numerical results for $\alpha_+(\chi)$, but this requires extending
the computation to much larger values of $\chi$ than those shown in
Fig.~\ref{fig:alpha-chi}.  Such a comparison shows a good agreement at
$\chi\gtrsim20$, see Fig.~\ref{fig:soliton-comparison}.

\subsection{Interpretation in terms of elementary excitations}
\label{sec:excitations}

In Sec.~\ref{sec:quadratic} we discussed the dispersions
(\ref{eq:KdV_dispersions}) of the elementary excitations of the
quantum KdV model (\ref{eq:H_kdv}) and demonstrated that they also
define the boundaries of the many-body spectrum of the system at
$L\to\infty$.  The elementary excitations were studied in
Ref.~\cite{pustilnik_fate_2015} with the help of the Bethe ansatz
solutions of the Lieb-Liniger \cite{lieb_exact_1963} and hyperbolic
Calogero-Sutherland \cite{sutherland_beautiful_2004} models, which
reduce to the quantum KdV model under proper limiting procedures.
Unfortunately, no mapping of this kind is known for the quantum mKdV
model (\ref{eq:H_mkdv}).  Furthermore, even though this model is
believed to be integrable \cite{sasaki_field_1987}, a Bethe ansatz
solution is not currently available.  Thus, we so far focused on
obtaining the boundaries of the many-body spectrum of the Hamiltonian
(\ref{eq:H_mkdv}) at $L\to\infty$.  We now discuss to what extent they
can be interpreted as elementary excitations of the system.

We start with the case of noninteracting fermions.  If the dispersion
is quadratic, the quasiparticle and quasihole with momentum $p$ have
energies $p^2/2m$ and $-p^2/2m$, respectively.  These values are the
highest and the lowest possible energies of any eigenstate with the
total momentum $p$.  In contrast, if the dispersion is cubic, both the
quasiparticle and quasihole have the same energy $\gamma p^3$, which
can be interpreted as a consequence of the particle-hole symmetry of
the Hamiltonian.  Due to the convexity of the dispersion $\gamma p^3$,
it also yields the upper boundary of the many-body spectrum.  The
lowest energy state can be constructed by distributing the total
momentum $p$ among a large number of quasiparticles and quasiholes
\cite{onefootnote}.  At $L\to\infty$, the corresponding energy is 0.

A similar interpretation of the boundaries of the spectrum applies in
the limit of strong attractive interactions, $\chi\to-\infty$.  In
this case, to leading order the Hamiltonian takes the form
(\ref{eq:H_mkdv_strong}).  The highest energy state has a single
bosonic excitation,
\begin{equation}
  \label{eq:single-boson_state_odd}
  \psi_o(p)=b_p^\dagger|0\rangle,
  \quad
  E_o(p)=\gamma\left(\frac12-\chi\right)p^3.
\end{equation}
Here the subscript `o' indicates that the state contains an odd number
of bosonic excitations, which means that it is odd with respect to
the particle-hole symmetry, see Eq.~(\ref{eq:particle-hole_boson}).
The highest energy even state is
\begin{equation}
  \label{eq:single-boson_state_even}
  \psi_e(p)=b_{p-p_1}^\dagger b_{p_1}^\dagger|0\rangle,
  \quad
  E_e(p)\simeq E_o(p)\left(1-\frac{3p_1}{p}\right),
\end{equation}
where $p_1=2\pi\hbar/L$.  In the limit of infinite system size, the
respective energies are equal, $E_o(p)=E_e(p)$.  The lowest energy
eigenstates of the Hamiltonian (\ref{eq:H_mkdv_strong}) with large
negative $\chi$ in the two parity sectors are
\begin{equation}
  \label{eq:multi-boson_state}
  \left(b_{p_1}^\dagger\right)^{\frac{p}{p_1}}|0\rangle,
  \quad
  \left(b_{p_1}^\dagger\right)^{\frac{p}{p_1}-2}b_{2p_1}^\dagger|0\rangle.
\end{equation}
The corresponding energies vanish at $L\to\infty$.

In the above examples the system has two branches of elementary
excitations, which correspond to states that are even and odd with
respect to particle-hole symmetry, with energies that become identical
at $L\to\infty$.  The dispersion $\varepsilon(p)$ of the excitations
defines the upper boundary of the excitation spectrum of the quantum
mKdV Hamiltonian (\ref{eq:H_mkdv}), while the lower boundary is at
zero energy.  Our results for the boundaries of the energy spectrum at
$L\to\infty$ given by Eq.~(\ref{eq:mkdv_boundaries_general}) and
Fig.~\ref{fig:alpha-chi} suggest that the same picture applies at all
$\chi<\chi_c\approx0.4$.  Then the dispersion of the elementary
excitations is
\begin{equation}
  \label{eq:upper_dispersion}
  \varepsilon_{\rm mKdV}^+(p)=\alpha_+(\chi)\gamma p^3.
\end{equation}
The function $\alpha_+(\chi)$ shown by the upper line in
Fig.~\ref{fig:alpha-chi} continues into the region $\chi>\chi_c$,
suggesting that the excitation branch with dispersion given by
Eq.~(\ref{eq:upper_dispersion}) exists at all interaction strengths.
On the other hand, the lower boundary of the spectrum corresponds to
negative energies at $\chi>\chi_c$.  Furthermore, at $\chi\to\infty$
the approximation (\ref{eq:H_mkdv_strong}) applies again.  The
dispersion associated with the two excitation branches
(\ref{eq:single-boson_state_odd}) and
(\ref{eq:single-boson_state_even}) is now concave and thus describes
the lower boundary of the energy spectrum.  This suggests that in
addition to the excitation branch (\ref{eq:upper_dispersion}), there
is another one, with dispersion
\begin{equation}
  \label{eq:lower_dispersion}
  \varepsilon_{\rm mKdV}^-(p)=\alpha_-(\chi)\gamma p^3,
  \quad
  \chi>\chi_c.
\end{equation}

Our results for the boundaries of the spectrum do not preclude the
possibility that this branch continues into the region $\chi<\chi_c$.
Indeed, a mode with energy $\alpha^*\gamma p^3$ would not affect the
boundaries of the many-body spectrum at $\chi<\chi_c$ as long as
$0<\alpha^*<\alpha_+$.  We note, however, that at $\chi=0$, when the
model describes free fermions, quasiparticles and quasiholes are the
only types of excitations present, and their dispersion is given by
Eq.~(\ref{eq:upper_dispersion}).  Thus the branch
(\ref{eq:lower_dispersion}) must terminate at some value of $\chi$
between 0 and $\chi_c$.  Some numerical evidence supporting this
scenario is obtained by studying overlaps of the single boson state
with all the eigenstates of the Hamiltonian with $0<\chi<\chi_c$
\cite{onefootnote}.

\section{Discussion of the results}
\label{sec:discussion}

In Sec.~\ref{sec:quadratic} we studied the elementary excitations and
the boundaries of the many-body spectrum of the system of spinless
chiral fermions with short-range interactions.  Upon bosonization the
problem reduces to the quantum KdV model (\ref{eq:H_kdv}), for which
exact results for the excitation spectrum are available
\cite{pustilnik_fate_2015}.  We concluded that at $p\ll p^*$ the
excitations are essentially the quasihole and quasiparticle of the
free Fermi gas, while at $p\gg p^*$, they are the bosonic density wave
and the KdV soliton.  Taking into consideration the definition
(\ref{eq:p^*}) of $p^*$, this can be alternatively interpreted as
crossover between the regimes of weak and strong interactions.  Given
the curvature of the dispersions of the two excitation branches,
Fig.~\ref{fig:dispersions_plot}, it is natural to expect that they
would coincide with the lowest and highest energies of many-body
states with total momentum $p$.  This conjecture is confirmed by our
numerical calculations.

Our main focus was on the study of the system of interacting spinless
chiral fermions with cubic dispersion, Sec.~\ref{sec:cubic}.
Bosonization reduces this problem to the quantum mKdV model
(\ref{eq:H_mkdv}).  The scaling properties of this model dictate that
the energy scales associated with it must be proportional to $p^3$.
This applies, in particular, to the boundaries of the many-body
spectrum, which we obtained numerically, see
Eq.~(\ref{eq:mkdv_boundaries_general}) and Fig.~\ref{fig:alpha-chi}.
Unlike the case of quadratic dispersion studied in
Sec.~\ref{sec:quadratic}, we found two distinct regimes of interaction
strength.  At $\chi>\chi_c$ the system behaves similarly to the
quantum KdV model (\ref{eq:H_kdv}) in that both boundaries of the
many-body spectrum can be viewed as states with one elementary
excitation.  The excitation belongs to one of two branches, with
dispersions given by Eqs.~(\ref{eq:upper_dispersion}) and
(\ref{eq:lower_dispersion}).  This analogy fails at $\chi<\chi_c$,
where only the upper boundary of the excitation spectrum behaves as an
elementary excitation; its dispersion is given by
Eq.~(\ref{eq:upper_dispersion}).  At $\chi=0$ and $\chi\to-\infty$ the
lower boundary corresponds to states with infinite number of
excitations, each carrying infinitesimal momentum.  It is natural to
apply the same interpretation to the lower boundary of the spectrum at
all $\chi<\chi_c$.

The existence of two qualitatively different regimes depending on the
interaction strength could have been anticipated by considering the
limit of strong interactions, $|\chi|\to\infty$.  In this limit the
system is described by the classical mKdV equation (\ref{eq:mkdv}).
The latter has harmonic wave solutions with infinitesimal amplitude,
for which the nonlinear term in Eq.~(\ref{eq:mkdv}) can be neglected.
They correspond to the bosonic excitations giving the upper (lower)
boundary of the spectrum at $\chi\to-\infty$ ($\chi\to+\infty$).  In
addition, at $\tilde\chi>0$ the mKdV equation (\ref{eq:mkdv}) has
soliton solutions.  In analogy with the quantum KdV model
(\ref{eq:H_kdv}), the soliton gives the upper boundary of the
excitation spectrum, see Sec.~\ref{sec:solitons}.  In contrast, no
solitons solutions exist at $\tilde\chi<0$, resulting in the
qualitatively different behavior of the lower boundary of the spectrum
at $\chi\to-\infty$.

\begin{figure}[t]
  \includegraphics[width=.45\columnwidth]{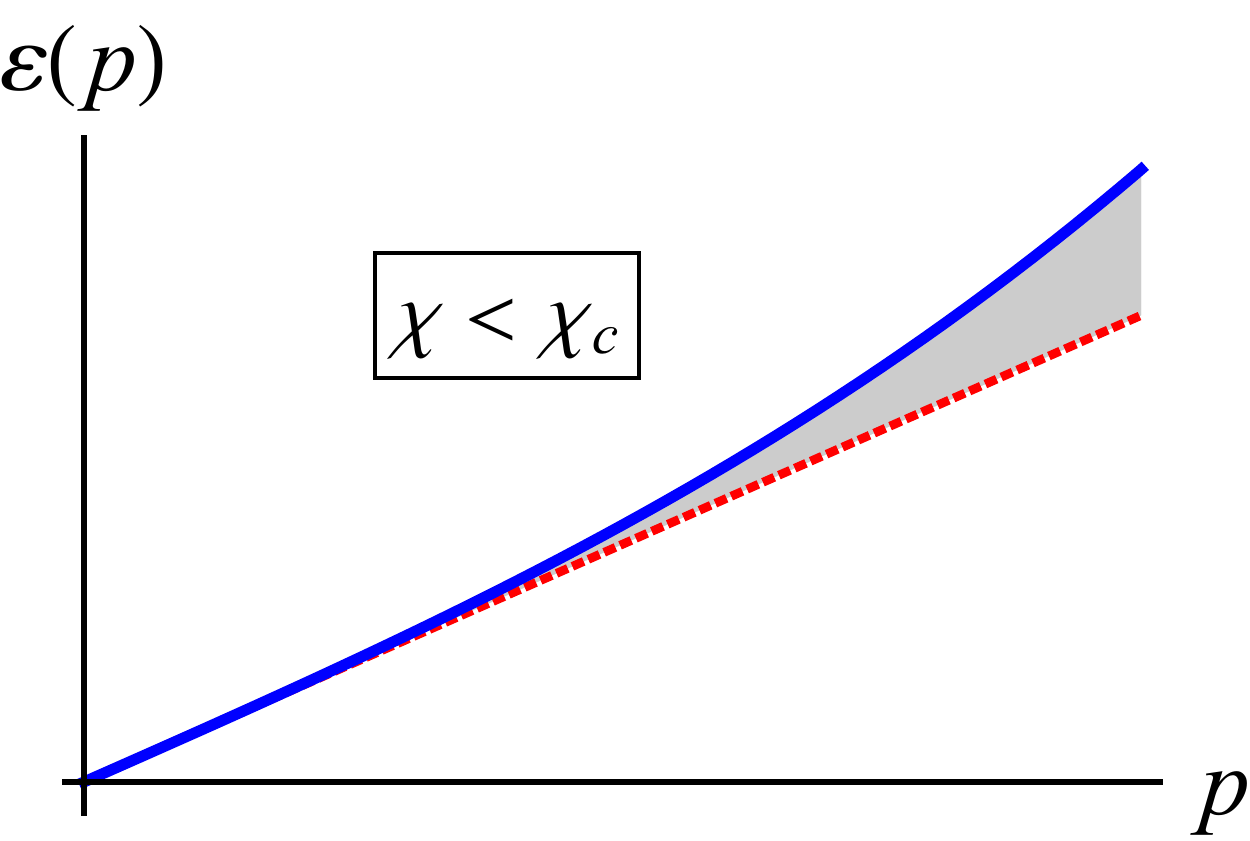}
  \quad
  \includegraphics[width=.45\columnwidth]{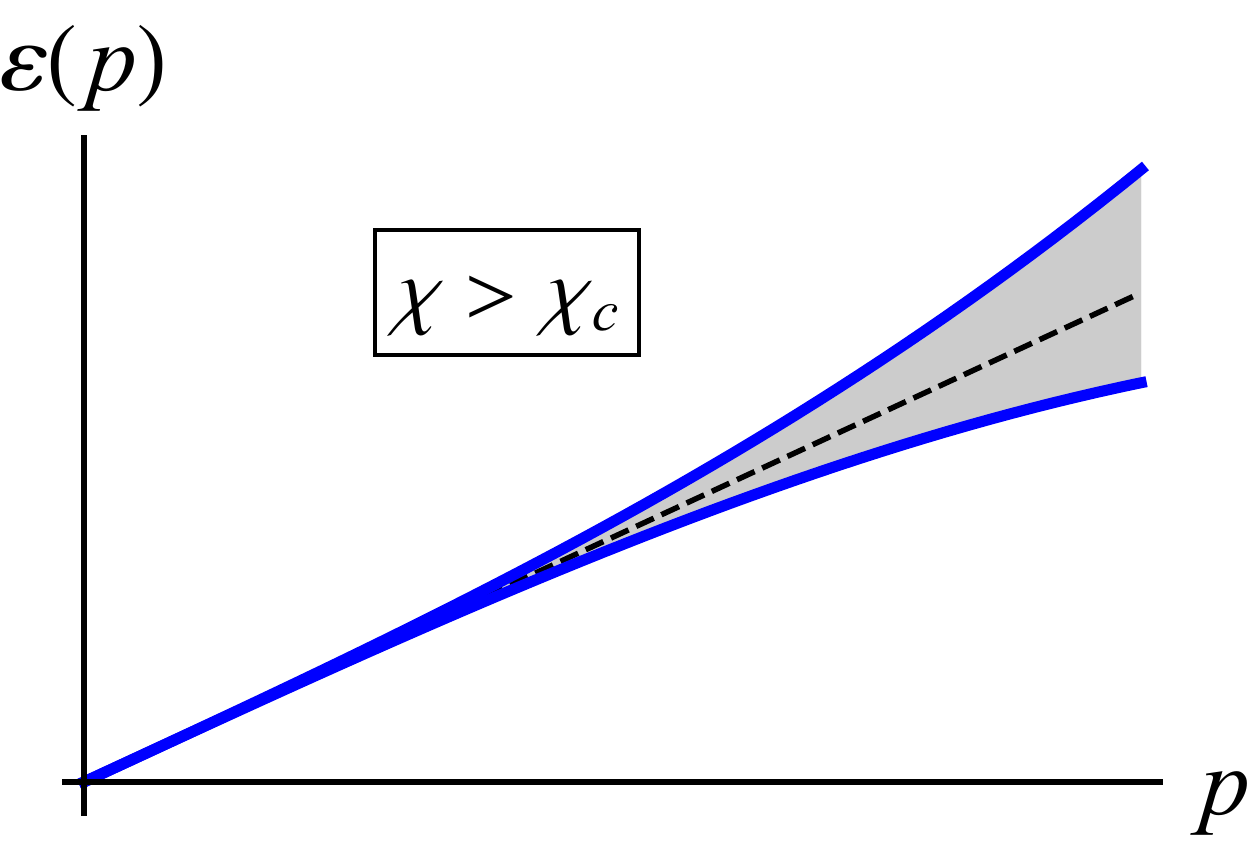}
  \caption{Boundaries of the many-body spectrum of the system of
    chiral fermions with cubic dispersion and short-range
    interactions.  For all $\chi$ the upper boundary is given by
    $\varepsilon=vp+\varepsilon_{\rm mKdV}^+(p)$, see
    Eq.~(\ref{eq:upper_dispersion}).  At $\chi<\chi_c$ the lower
    boundary of the spectrum is the straight dotted line
    $\varepsilon=vp$, while at $\chi>\chi_c$ it is given by
    $\varepsilon=vp+\varepsilon_{\rm mKdV}^-(p)$, see
    Eq.~(\ref{eq:lower_dispersion}).}
\label{fig:mkdv-dispersions}
\end{figure}

The energy eigenvalues of the original model of chiral fermions with
cubic dispersion are related to those of the quantum mKdV model by
Eq.~(\ref{eq:energy_relation_mkdv}).  Our results are illustrated in
Fig.~\ref{fig:mkdv-dispersions}, where the boundaries of the energy
spectrum are shown in the two regimes, $\chi<\chi_c$ and
$\chi>\chi_c$.  The behavior of the energy spectrum in the latter
regime is similar to that for fermions with quadratic dispersion, see
Fig.~\ref{fig:dispersions_plot}.

Let us now briefly discuss the expected behavior of the dynamic
response functions of the system, which include the spectral function
$A(p,\epsilon)$ and the dynamic structure factor $S(p,\epsilon)$.  The
shaded regions in Figs.~\ref{fig:dispersions_plot} and
\ref{fig:mkdv-dispersions} correspond to the possible energies of the
system at a given momentum.  The response functions must vanish
outside these regions.  In nonchiral systems the behavior of
$A(p,\epsilon)$ and $S(p,\epsilon)$ near the edge of support was
studied phenomenologically using the mobile impurity model
\cite{imambekov_phenomenology_2009, imambekov_one-dimensional_2012}.
This approach should be applicable near the boundaries of the spectrum
of chiral fermions with quadratic dispersion, which allow for the
interpretation as states with one elementary excitation.  It should
also apply to the boundaries shown by solid lines in
Fig.~\ref{fig:mkdv-dispersions} for the case of fermions with cubic
dispersion.  In the nonchiral case both $A(p,\epsilon)$ and
$S(p,\epsilon)$ were predicted to scale as a power of the distance
from the boundary \cite{imambekov_phenomenology_2009,
  imambekov_one-dimensional_2012}.  We expect an analogous power-law
scaling in the chiral case.

The lower boundary of the spectrum at $\chi<\chi_c$, shown by the
dotted line in Fig.~\ref{fig:mkdv-dispersions}, has a different
nature.  The states near this boundary involve a large number of
excitations with very small momenta.  A similar problem has been
studied in the case of phonons in liquid helium, where the exponential
suppression of the response was found
\cite{iordanskii_properties_1978}.  In the case of one-dimensional
systems with linear spectrum, equivalent to the limit $\chi\to-\infty$
of our model, the exponential suppression of the spectral function was
found in Ref.~\cite{matveev_spectral_2022}.  Furthermore, in the case
of weakly interacting fermions, $|\chi|\ll1$, the overlap of a
low-energy state involving a large number of particle-hole pairs with
that involving a single pair occurs in a high order of the
perturbation theory.  As a result, the response functions must again
be exponentially small near the boundary.  We therefore expect
exponential suppression of the dynamic response functions near the
lower boundary of the spectrum at $\chi<\chi_c$.  Our numerical
treatment of the Hamiltonian (\ref{eq:H_mkdv}) supports this
conclusion \cite{onefootnote}.

The qualitative change of the energy spectrum at the interaction
strength $\chi=\chi_c$ can be interpreted as a phase transition in the
system.  Interestingly, this phase transition is purely dynamic, in
that it appears only in the dynamic response functions of the system.
Indeed, due to the chiral nature of the problem, the ground state
$|0\rangle$, which corresponds to the filled Fermi surface, does not
depend on the interaction strength.  Thus the static properties of the
system are not affected by the interactions.  Alternatively, one can
consider the behavior of the system at a fixed value $p$ of the total
momentum.  In this case the ground state energy as a function of the
interaction strength $\chi$ shows nonanalytic behavior at
$\chi=\chi_c$, which can be interpreted as a quantum phase transition.
A similar nonanalytic behavior of the boundary of the energy spectrum
at a fixed momentum was recently found in the system of chiral
fermions with quadratic spectrum and Coulomb interactions
\cite{martin_scar_2022}.  Finally, we note that in our numerical data
represented in Fig.~\ref{fig:alpha-chi} we have not been able to find a
deviation of $\chi_c$ from 0.4, i.e., we expect that the exact value
is $\chi_c=\frac25$.

\begin{acknowledgments}

  The author is grateful to A.~Furusaki, I.~Martin, M.~R.~Norman, and
  M.~Pustilnik for helpful discussions.  This work was supported by
  the U.S. Department of Energy, Office of Science, Basic Energy
  Sciences, Materials Sciences and Engineering Division.
  
\end{acknowledgments}

\onecolumngrid
\newpage
\setcounter{equation}{0}
\setcounter{figure}{0}
\setcounter{section}{0}

\renewcommand{\theequation}{S\arabic{equation}}
\renewcommand{\thepage}{S\arabic{page}}
\renewcommand{\thesection}{S\arabic{section}}
\renewcommand{\thetable}{S\arabic{table}}
\renewcommand{\thefigure}{S\arabic{figure}}

\renewcommand{\bibnumfmt}[1]{[{\normalfont S#1}]}


\begin{center}
	{\large\textbf{\TITLE}
		\\\vskip 5pt
		\normalsize{--Supplemental Material--}
		\\\vskip 5pt
	}
	
        K. A. Matveev \vskip 0.5mm \textit{\small Materials Science
          Division, Argonne National Laboratory, Argonne, Illinois
          60439, USA}
	
\end{center}
\vskip 1.5pt
\twocolumngrid

\setcounter{page}{1}

\section{Numerical treatment of the quantum KdV Hamiltonian
  (\ref{eq:H_kdv})}

\label{sec:numerics_KdV}

We start by setting $N=0$ and choosing the units in which $\hbar=m=1$
and $L=2\pi$.  We then use the expression (\ref{eq:phi}) for $\phi(x)$
to rewrite the Hamiltonian (\ref{eq:H_kdv}) in terms of the bosonic
operators $b_l^{}$ and $b_l^\dagger$,
\begin{eqnarray}
  \label{eq:H_kdv_b}
  H_{\rm KdV}
  &=&\frac12\sum_{l=2}^\infty \sum_{l'=1}^{l-1}
  \sqrt{ll'(l-l')}
  \left(b_l^\dagger b_{l'}^{}b_{l-l'}^{}+ b_{l'}^\dagger
    b_{l-l'}^\dagger b_l^{}\right)
\nonumber\\
  &&-\frac{\eta}{2\pi}\sum_{l=1}^\infty
  l^3b_l^\dagger b_l^{}.
\end{eqnarray}
We now fix the total momentum $P$ of the system and construct the
basis of the Hilbert space by partitioning momentum as a sum
$P=\sum_lln_l$ over all bosonic states $l$ with occupation numbers
$n_l$.  Each integer partition of $P$ defines a unique basis state.
In this basis the Hamiltonian (\ref{eq:H_kdv_b}) is a real symmetric
matrix.  The dimension of the Hilbert space is number of integer
partitions of $P$.  We perform subsequent calculations for $P$ up to
48, for which there are 147273 integer partitions.

\begin{figure}[b]
\includegraphics[width=.45\textwidth]{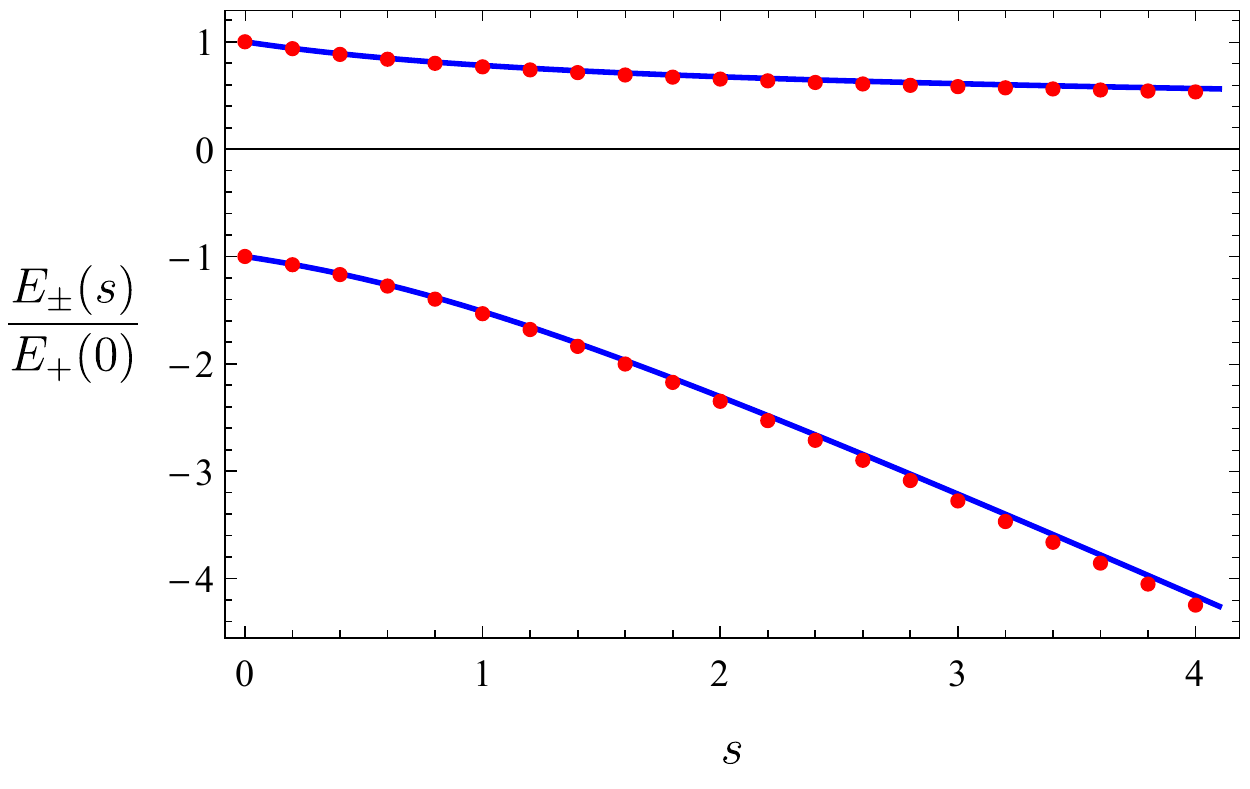}
\caption{Comparison of the quasiparticle energies
  (\ref{eq:KdV_dispersions}) with the spectral boundaries of the
  Hamiltonian (\ref{eq:H_kdv}).  The solid lines are $e_\pm(s)/s^2$,
  cf.~Fig.~\ref{fig:KdV-numerics}.  The dots represent the numerically
  computed boundaries of the energy spectrum $E_\pm(s)$ of the Hamiltonian
  (\ref{eq:H_kdv}) for $P=48$, rescaled by $E_+(0)$.}
\label{fig:KdV-numerics-48}
\end{figure}

Next, we find the highest and the lowest eigenstates of the
Hamiltonian, $E_+$ and $E_-$, numerically for several values of the
interaction constant $\eta$.  It is convenient to present the results
in terms of the parameter $s=p/p^*=pm\eta/\pi\hbar$, which in our
units becomes $s=\eta P/\pi$.  In Fig.~\ref{fig:KdV-numerics-48} we plot
$E_\pm(s)$ normalized by $E_+(0)=P(P-1)/2$, which ensures that the ratio
is $\pm 1$ at $s=0$.

Numerical results for $E_\pm(s)/E_+(0)$ are compared with
$e_\pm(s)/s^2$, which is the expected result for the system of
infinite size, see Sec.~\ref{sec:quadratic}.  In our calculation the
dimensionless momentum is $P=pL/2\pi\hbar$.  Thus the limit
$L\to\infty$ at fixed $p$ corresponds to $P\to\infty$.  The data for
$P=48$, see Fig.~\ref{fig:KdV-numerics-48}, shows good, but not
perfect agreement, especially at larger values of $s$.

\begin{figure}[b]
\includegraphics[width=.45\textwidth]{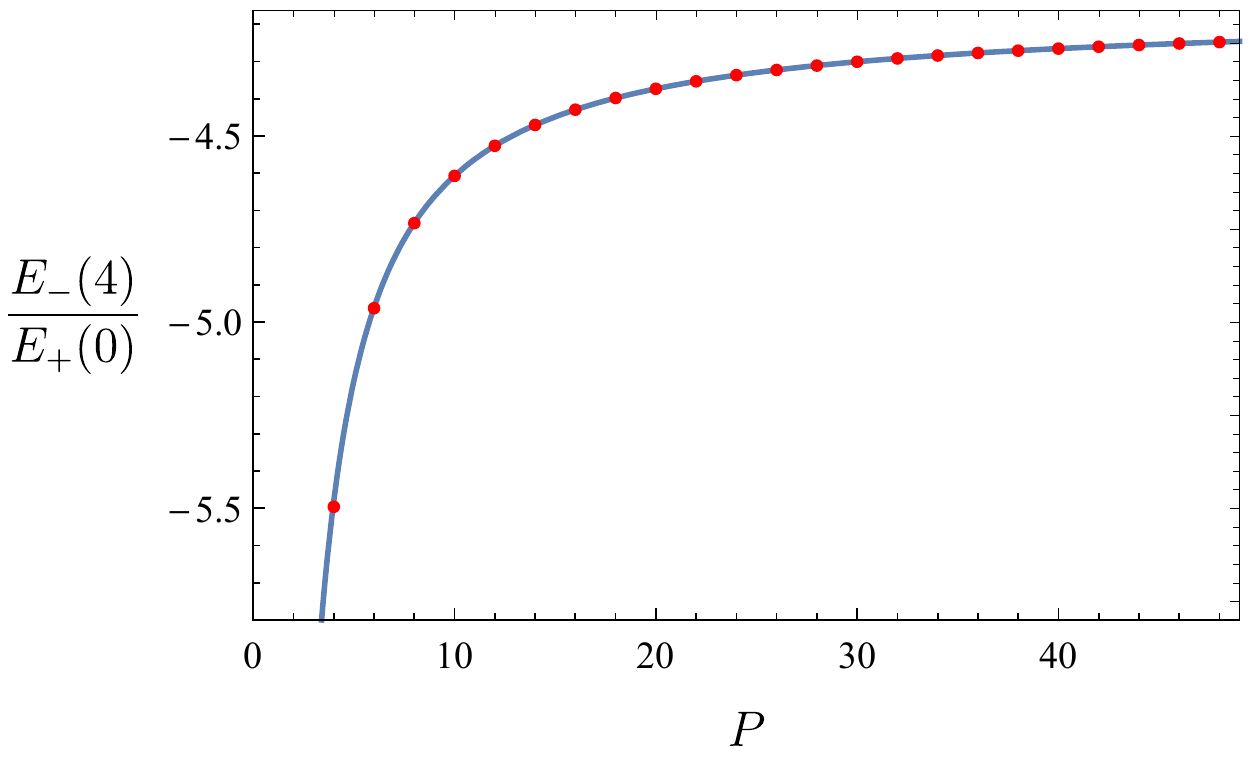}
\caption{The ratio $E_-(4)/E_+(0)$ evaluated for the even values of
  the total momentum $P$ from 4 to 48.  Solid line is the fit of the
  form (\ref{eq:fitfunction}) using four data points $P=42$, 44, 46, and 48.
}
\label{fig:numerical_fits}
\end{figure}

Since the dimension of the Hilbert space grows exponentially with $P$,
we cannot increase $P$ indefinitely.  On the other hand, our data can
be reliably extrapolated to $P\to\infty$.  To illustrate the procedure
we show in Fig.~\ref{fig:numerical_fits} the computed values of
$E_\pm(4)/E_+(0)$ (bottom right red dot in
Fig.~\ref{fig:KdV-numerics-48}) for even values of $P$ between 4 and
48.  We then fit the data points for $P=42$, 44, 46, and 48 to the
function
\begin{equation}
  \label{eq:fitfunction}
  f(P)=A_0+\frac{A_1}{P}+\frac{A_2}{P^2}+\frac{A_3}{P^3}
\end{equation}
and find the coefficients $A_0$, $A_1$, $A_2$, and $A_3$.  The
resulting function $f(P)$ is shown by the solid line in
Fig.~\ref{fig:numerical_fits}.  Given the excellent quality of the
fit, we conclude that replacing the computed value of
$E_\pm(4)/E_+(0)$ for $P=48$ with $f(\infty)=A_0$ will accurately
extrapolate the results to infinite system size.  We perform this
fitting procedure for all the points shown in
Fig.~\ref{fig:KdV-numerics-48}.  The resulting data is shown in
Fig.~\ref{fig:KdV-numerics}.

\pagebreak
\begin{widetext}
  \section{Numerical treatment of the quantum mKdV Hamiltonian
  (\ref{eq:H_mkdv})}

\label{sec:numerics_mKdV}

We again set $N=0$ and choose the units so that $\hbar=\gamma=1$ and
$L=2\pi$.  Substitution of Eq.~(\ref{eq:phi}) into the Hamiltonian
(\ref{eq:H_mkdv}) yields

\begin{eqnarray}
  \label{eq:H_mkdv_b}
  H_{\rm mKdV}
  &=&
  \sum_{l=1}^\infty\sum_{l'=1}^\infty\sum_{l''=1}^\infty
  \sqrt{ll'l''(l+l'+l'')}
  \big(b_l^\dagger b_{l'}^\dagger b_{l''}^\dagger b_{l+l'+l''}^{}
  +b_{l+l'+l''}^\dagger b_l^{} b_{l'}^{} b_{l''}^{}\big)
\nonumber\\
  &&+\frac32\sum_{l=2}^\infty\sum_{l'=1}^{l-1}\sum_{l''=1}^{l-1}
     \sqrt{l'(l-l')l''(l-l'')}
     b_{l'}^\dagger b_{l-l'}^\dagger b_{l''}^{} b_{l-l''}^{}
     +\left(\frac12-\chi\right)
     \sum_{l=1}^\infty l^3 b_l^\dagger b_l^{}.
\end{eqnarray}
\end{widetext}
To study the energy spectrum of this Hamiltonian at a fixed value of
the total momentum $P$, we use the approach described in
Sec.~\ref{sec:numerics_KdV}, see the comments below
Eq.~(\ref{eq:H_kdv_b}).

\subsection{Eigenvalues}
\label{sec:eigenvalues}

In Fig.~\ref{fig:all-energies} we show all 22 eigenvalues of the
Hamiltonian $H_{\rm mKdV}-\frac14\hat P$ for $P=8$ and the interaction
strength paramenter in the range $-2\leq\chi\leq2$.  Compared to the
operator $H_{\rm mKdV}$, for which the relevant energy scales are
proportional to $P^3$, the term $-\hat P/4$ is small and amounts to a
finite-size correction.  The origin of this term is the contribution
$-(\pi^2\hbar^2\gamma/L^2)\hat P$ in Eq.~(\ref{eq:H_cubic}).  In the
case of zero interaction, $\chi=0$, the operator
$H_{\rm mKdV}-\frac14\hat P$ is the exact bosonized form of the
operator
\begin{equation}
  \label{eq:H_3}
  H_3=\sum_p\left(p-\frac12\right)^3 a_p^\dagger a_p^{},
\end{equation}
where the momentum $p$ in our units takes integer values; cf.\
Eqs.~(\ref{eq:H_general}) and (\ref{eq:cubic_nonlinearity}).  At a
given total momentum $P$ the maximum eigenvalues of $H_3$ correspond
to a single fermion moved from the highest occupied state $p=0$ to
the state with $p=P$ (quasiparticle), or from $p=1-P$ to $p=1$
(quasihole).  In both cases the energy is
\begin{equation}
  \label{eq:E^max}
  E_3^{\rm max}(P)=\left(P-\frac12\right)^3+\frac{1}{2^3}
\end{equation}
In the case of $P=8$ we have $E_3^{\rm max}(8)=422$ which agrees with
the maximum eigenvalue at $\chi=0$ in Fig.~\ref{fig:all-energies}.
The full set of the 22 numerically obtained eigenvalues of
$H_{\rm mKdV}-\frac14\hat P$ at $\chi=0$ is
\begin{eqnarray*}
  E&=&422., 422., 278., 278., 182., 182., 170., 170., 134., 134.,\\
  &&107., 107., 98., 98., 86., 62., 62., 62., 62., 62., 62., 38.
\end{eqnarray*}
It is easy to check directly that they are exactly the eigenvalues of
the Hamiltonian (\ref{eq:H_3}) at total momentum $P=8$.

\subsection{Evaluation of $\alpha_\pm(\chi)$}
\label{sec:alphas}

\begin{figure}[b]
\includegraphics[width=.45\textwidth]{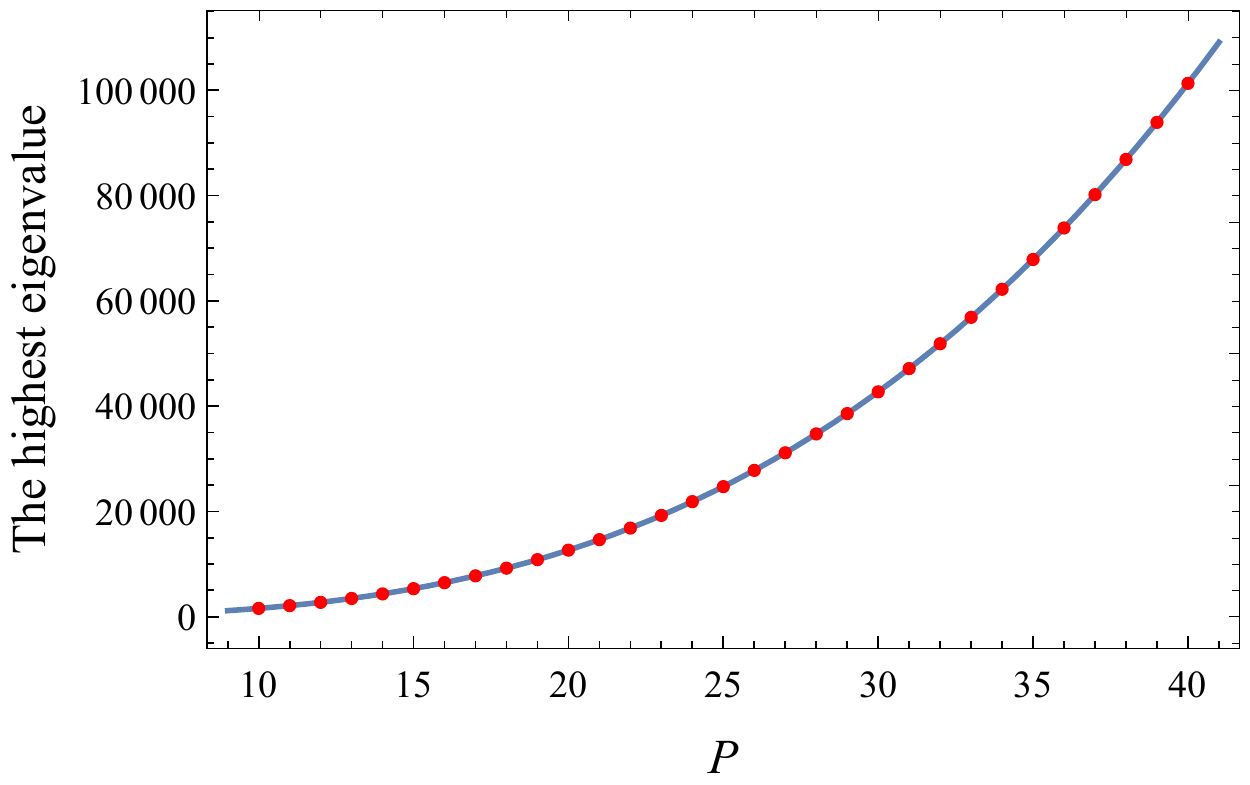}
\caption{The dots represent the numerically obtained results for the
  highest eigenvalue of the Hamiltonian $H_{\rm mKdV}-\frac14\hat P$
  for $\chi=-1$ and $P=10,11,\ldots,40$.  The solid line is the cubic
  polynomial $1.5849 P^3-0.019308 P^2-2.5392 P+12.884$ obtained by
  fitting the the data for $P=37$, 38, 39, and 40.  The fit yields
  $\alpha_+(-1)\approx1.585$, cf.~Fig.~\ref{fig:alpha-chi}.}
\label{fig:fit_alpha_+}
\end{figure}

To obtain the results for $\alpha_+(\chi)$ shown by the top line in
Fig.~\ref{fig:alpha-chi}, we obtain the highest eigenvalue of the
Hamiltonian $H_{\rm mKdV}-\frac14\hat P$ with a given $\chi$ for $P$
between 10 and 40.  We then fit the data for the four highest values
of $P$ to a cubic polynomial
\[
  B_0 + B_1 P+ B_2 P^2+ B_3 P^3
\]
and identify $\alpha_+(\chi)$ as the coefficient $B_3$.  (Since in
physical units $P=pL/2\pi\hbar$, extrapolation to $P\to\infty$ is
equivalent to taking the limit $L\to\infty$ at fixed physical momentum
$p$.)  The fits are excellent, see Fig.~\ref{fig:fit_alpha_+}; the
results for $\alpha_+(\chi)$ are shown by the upper line in
Fig.~\ref{fig:alpha-chi}.  The same method is used to obtain the
results for $\alpha_-(\chi)$ shown by the lower line in
Fig.~\ref{fig:alpha-chi} at $\chi>0.4$.

The fitting procedure described above cannot be applied to the
evaluation of $\alpha_-(\chi)$ at $\chi<0.4$.  To illustrate the
difficulty, we consider the case $\chi=0$, which corresponds to the
free fermion Hamiltonian (\ref{eq:H_3}).  If $P=n^2$, where $n$ is
integer, the eigenstate with the lowest energy is obtained from
$|0\rangle$ by moving fermions from $n$ highest occupied states to $n$
lowest empty states,
\[
  a_{n}^\dagger\ldots a_2^\dagger a_1^\dagger
  a_0^{}a_{-1}^{}\ldots a_{-(n-1)}^{}|0\rangle.
\]
The corresponding energy is $(2n^4-n^2)/4=(2p^2-p)/4$.  If
$n^2<p<(n+1)^2$, the total number of displaced fermions remains $n$.
The extra momentum $p-n^2$ is accommodated by moving some of the newly
created particles above the Fermi level up in momentum by 1 and/or
some of the new holes down by 1.  The resulting lowest eigenvalue is 
\begin{widetext}
\begin{equation}
  \label{eq:E^min}
  E_3^{\rm min}(P)=
  \left\{
    \begin{array}[c]{ll}
      \frac14\left(2P^2-P\right),& P=n^2,
      \\[1ex]
      \frac{1}{8} \left(4 n^4+8 n^3+10n^2+6 n+1
      -\left(2 n^2+2 n+1-2 P\right)^3\right), & n^2<P\leq n^2+n,
      \\[1ex]
      \frac{1}{8} \left(4 n^4+16 n^3+22n^2+12 n+1
      -\left(2 n^2+4 n+1-2 P\right)^3\right), & n^2+n<P\leq n^2+2n,
      \end{array}
      \right.
    \end{equation}
\end{widetext}
where $n$ is the integer part of $\sqrt{P}$.  It is easy to see that
$E_3^{\rm min}(P)$ grows as $P^2/2$ at $P\to\infty$.  This yields
\[
  \alpha_-(0)= \lim_{P\to\infty}\frac{E_3^{\rm min}(P)}{P^3}=0.
\]

\begin{figure}
\includegraphics[width=.45\textwidth]{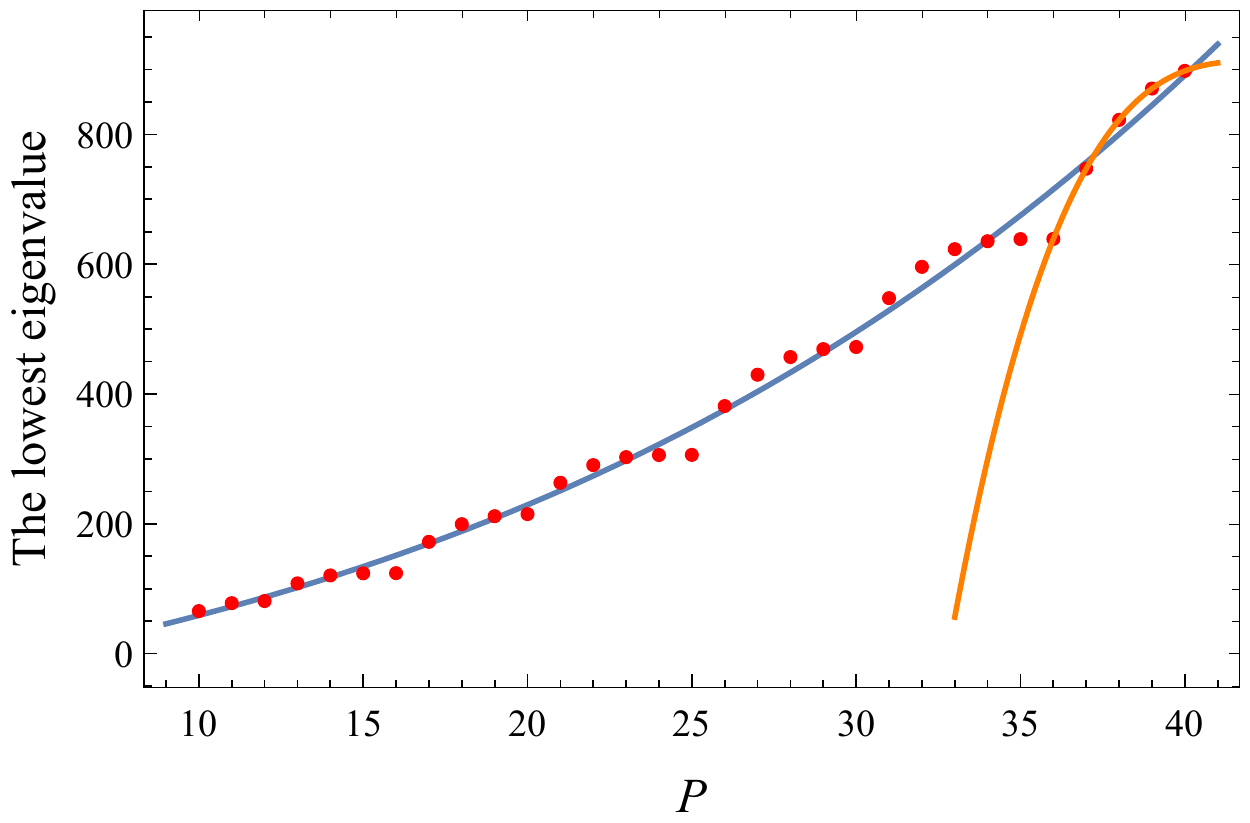}
\caption{The lowest eigenvalue $E_3^{\rm min}(P)$ of the Hamiltonian
  (\ref{eq:H_3}) for the total momentum $P=10$, 11, \ldots, 40.  The
  blue line is the best fit of the data to a cubic polynomial.  The
  orange line is the same fit using only the last four data points.}
\label{fig:free-fermion-fits}
\end{figure}

We show $E_3^{\rm min}(P)$ for $P$ between 10 and 40 in
Fig.~\ref{fig:free-fermion-fits}.  Clearly, fitting the last four data
points ($P=37$, 38, 39, 40) yields a wrong trend at large $P$.  On the
other hand, a fit using all the data points in
Fig.~\ref{fig:free-fermion-fits}, yields
$0.00547362 P^3+0.151921 P^2+8.64975 P-48.2719$, corresponding to
$\alpha_-(0)=0.00547362\ll1$.  This small value of $\alpha_-(0)$ is a
reasonable approximation to the exact analytical result
$\alpha_-(0)=0$.

We apply this fitting procedure to the lowest energy eigenvalues for
all $\chi<0.4$ and find $|\alpha_-(\chi)|\leq\alpha_-(0)$ in the whole
range $-2<\chi<0.4$.  We thus conclude that to the numerical accuracy,
$\alpha_-(\chi)=0$ for all $\chi<\chi_c\approx 0.4$, as shown in
Fig.~\ref{fig:alpha-chi}.
    
\subsection{Overlap of some eigenstates}
\label{sec:overlaps}

In Sec.~\ref{sec:discussion} we argued that the behavior of the
dynamic response functions at the lower boundary of the spectrum is
qualitatively different in the regimes $\chi<\chi_c$ and
$\chi>\chi_c$.  Specifically, both the spectral function and the
dynamic structure factor are expected to show power-law dependence on
the distance from the lower boundary at $\chi>\chi_c$ and be
exponentially suppressed at $\chi<\chi_c$.  Here, we consider the
dynamic structure factor, which in bosonic variables is given by
\begin{equation}
  \label{eq:DSF}
  S_p(\epsilon)=p\sum_j|\langle j|b_P^\dagger|0\rangle|^2
  \delta(\epsilon-E_j),
\end{equation}
where $p=(2\pi\hbar/L)P$, the summation is over all the eigenstates
$|j\rangle$ of the Hamiltonian (\ref{eq:H_mkdv_b}), and the $E_j$ is
the energy of the state $|j\rangle$ measured from that of the vacuum
state $|0\rangle$.

\begin{figure}
\includegraphics[width=.45\textwidth]{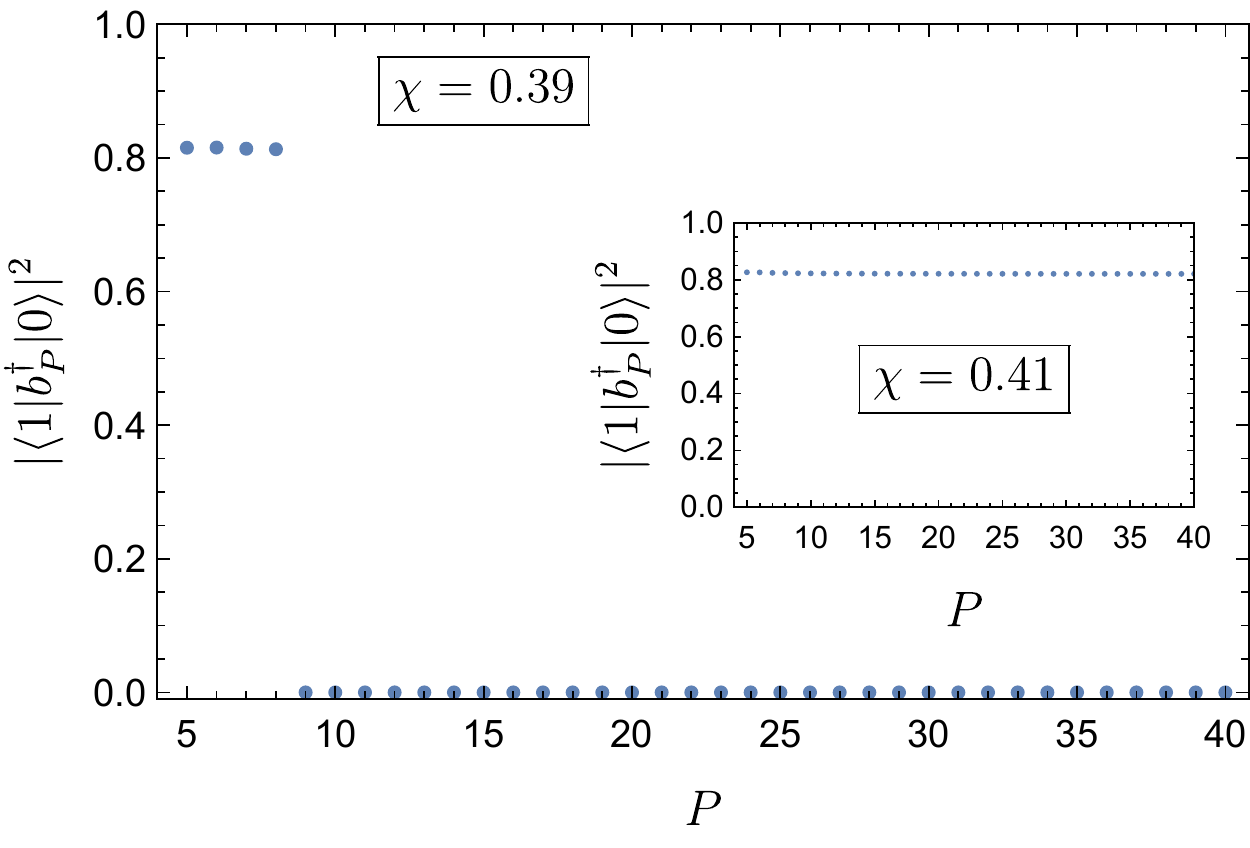}
\caption{Dependence of $|\langle 1|b_P^\dagger|0\rangle|^2$ on $P$ for
  $\chi=0.39<\chi_c$ and $P$ from 5 to 40.  At $P\geq9$ the overlap is
  very small.  In contrast, at $\chi=0.41>\chi_c$ (inset) the overlap is close
  to $0.8$ for all $P$.}
\label{fig:overlaps}
\end{figure}

The exact diagonalization of the Hamiltonian (\ref{eq:H_mkdv_b})
yields both the eigenstates and the overlaps
$\langle j|b_P^\dagger|0\rangle$.  However, numerical evaluation of
$S_p(\epsilon)$ in the limit $L\to\infty$ using Eq.~(\ref{eq:DSF}) is
not straightforward, as the dimension of the Hilbert space grows
exponentially with $P$.  Our argument is Sec.~\ref{sec:discussion} was
that at $\chi<\chi_c$ the states near the lower boundary of the
spectrum involve a large number of quasiparticles, which should result
in an exponentially small overlap with the state
$b_P^\dagger|0\rangle$.  This can be demonstrated numerically.  In
Fig.~\ref{fig:overlaps} we show the dependence of
$|\langle 1|b_P^\dagger|0\rangle|^2$ on $P$, where $|1\rangle$ is the
lowest energy eigenstate of the Hamiltonian (\ref{eq:H_mkdv_b}).  For
$\chi=0.39$, which is slightly below the crossover value
$\chi_c\approx 0.4$, the overlap drops sharply with increasing $P$.  A
closer examination of the data shows an exponential dependence of the
overlap on $P$.  At $P=40$ we obtained
$|\langle 1|b_P^\dagger|0\rangle|^2\approx 4\times 10^{-30}$.  A
similar behavior is observed for other values of $\chi$ below
$\chi_c$.  On the other hand, for $\chi>\chi_c$ the overlap shows only
a weak dependence on $P$.

\begin{figure}[b]
\includegraphics[width=.45\textwidth]{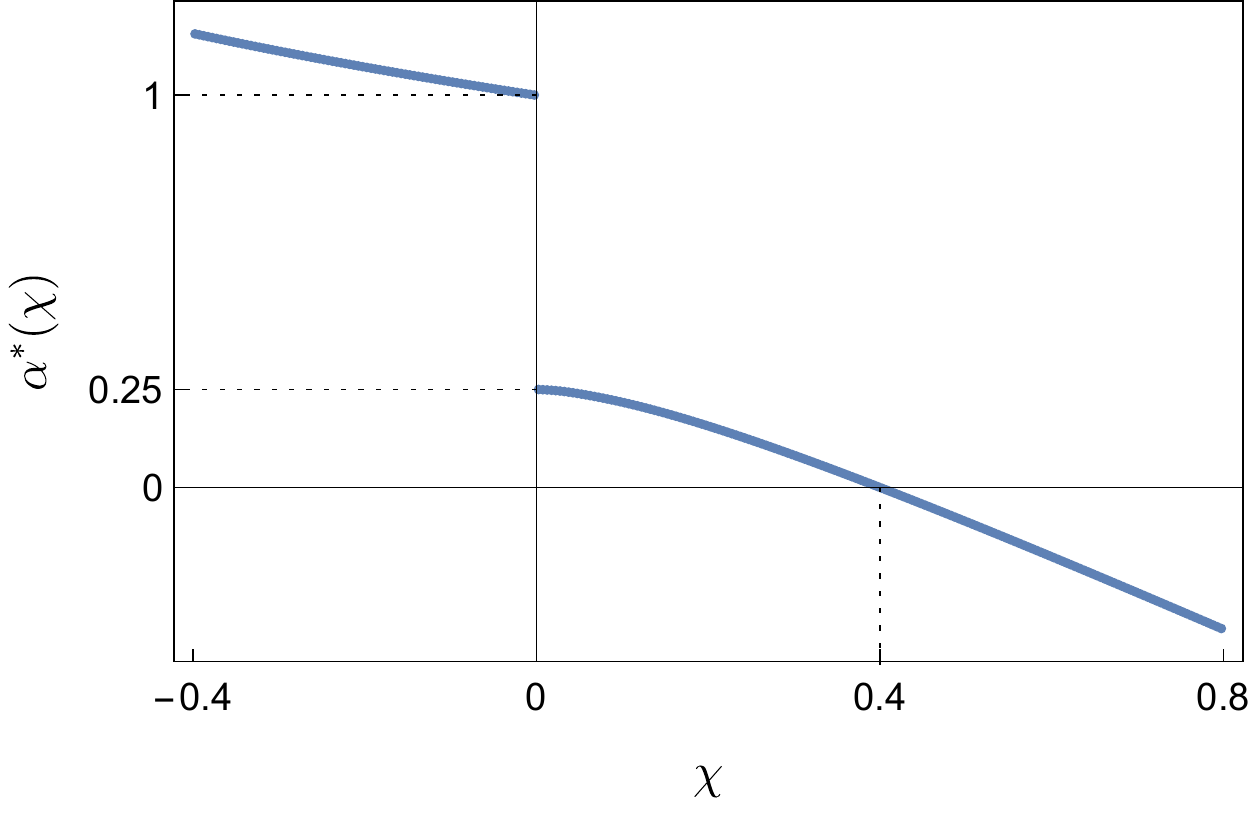}
\caption{Function $\alpha^*(\chi)$ that parametrizes the energy
  $\alpha^*P^3$ of the eigenstate $|j\rangle$ that has the
  largest overlap with the single boson state $b_P^\dagger|0\rangle$.}
\label{fig:maxoverlaps}
\end{figure}

At the end of Sec.~\ref{sec:cubic} [below
Eq.~(\ref{eq:lower_dispersion})] we mentioned the possibility of the
second branch of elementary excitations existing at $0<\chi<\chi_c$.
The energy of this mode $\alpha^*\gamma p^3$ would have to be positive
but lower than the energy of the first branch
(\ref{eq:upper_dispersion}), i.e., $0<\alpha^*<\alpha_+$.  To explore
this possibility, we evaluated the overlaps of the single boson state
$b_P^\dagger|0\rangle$ with all the eigenstates $|j\rangle$ of the
Hamiltonian (\ref{eq:H_mkdv_b}) for various values of $P$ and $\chi$.
For a given $\chi$, we then identified the state with the largest
overlap $|\langle j|b_P^\dagger|0\rangle|^2$ for each $P$ and obtained
the corresponding energy $E^*(P)$.  After numerical extrapolation to
$P\to\infty$, we obtained $\alpha^*=\lim_{P\to\infty}E^*(P)/P^3$.  The
dependence $\alpha^*(\chi)$ is shown in Fig.~\ref{fig:maxoverlaps}.

We note that at $\chi<0$ the state with the largest overlap with
$b_P^\dagger|0\rangle$ is the highest energy state.  Thus,
\begin{equation}
  \label{eq:alpha-star-neg-chi}
  \alpha^*(\chi)=\alpha_+(\chi),
  \quad
  \chi<0.
\end{equation}
In other words, at negative $\chi$ our procedure has identified the
previously obtained branch of elementary excitations
(\ref{eq:upper_dispersion}).  On the other hand, at
$\chi>\chi_c\approx0.4$ the maximum overlap is achieved for the ground
state of the system, i.e.,
\begin{equation}
  \label{eq:alpha-star-chi-gtr-chi_c}
  \alpha^*(\chi)=\alpha_-(\chi),
  \quad
  \chi>\chi_c.
\end{equation}
In this case our procedure yields the lower of the two branches of
elementary excitations, Eq.~(\ref{eq:lower_dispersion}).  In the
limits $\chi\to\pm\infty$ the results (\ref{eq:alpha-star-neg-chi})
and (\ref{eq:alpha-star-chi-gtr-chi_c}) were expected, because in the
interaction-dominated regimes bosons are elementary excitations of the
system.

The data shown in Fig.~\ref{fig:maxoverlaps} suggests that the lower
branch (\ref{eq:lower_dispersion}) of the elementary excitations at
$\chi>\chi_c$ continues into the region $0<\chi<\chi_c$, where its
energy is positive.  The latter property, corresponding to
$\alpha^*>0$ in Fig.~\ref{fig:maxoverlaps}, means that this branch is
no longer associated with the ground state of the system.



\begin{thebibliography}{23}%
\makeatletter
\providecommand \@ifxundefined [1]{%
 \@ifx{#1\undefined}
}%
\providecommand \@ifnum [1]{%
 \ifnum #1\expandafter \@firstoftwo
 \else \expandafter \@secondoftwo
 \fi
}%
\providecommand \@ifx [1]{%
 \ifx #1\expandafter \@firstoftwo
 \else \expandafter \@secondoftwo
 \fi
}%
\providecommand \natexlab [1]{#1}%
\providecommand \enquote  [1]{``#1''}%
\providecommand \bibnamefont  [1]{#1}%
\providecommand \bibfnamefont [1]{#1}%
\providecommand \citenamefont [1]{#1}%
\providecommand \href@noop [0]{\@secondoftwo}%
\providecommand \href [0]{\begingroup \@sanitize@url \@href}%
\providecommand \@href[1]{\@@startlink{#1}\@@href}%
\providecommand \@@href[1]{\endgroup#1\@@endlink}%
\providecommand \@sanitize@url [0]{\catcode `\\12\catcode `\$12\catcode
  `\&12\catcode `\#12\catcode `\^12\catcode `\_12\catcode `\%12\relax}%
\providecommand \@@startlink[1]{}%
\providecommand \@@endlink[0]{}%
\providecommand \url  [0]{\begingroup\@sanitize@url \@url }%
\providecommand \@url [1]{\endgroup\@href {#1}{\urlprefix }}%
\providecommand \urlprefix  [0]{URL }%
\providecommand \Eprint [0]{\href }%
\providecommand \doibase [0]{https://doi.org/}%
\providecommand \selectlanguage [0]{\@gobble}%
\providecommand \bibinfo  [0]{\@secondoftwo}%
\providecommand \bibfield  [0]{\@secondoftwo}%
\providecommand \translation [1]{[#1]}%
\providecommand \BibitemOpen [0]{}%
\providecommand \bibitemStop [0]{}%
\providecommand \bibitemNoStop [0]{.\EOS\space}%
\providecommand \EOS [0]{\spacefactor3000\relax}%
\providecommand \BibitemShut  [1]{\csname bibitem#1\endcsname}%
\let\auto@bib@innerbib\@empty
\bibitem [{\citenamefont {Lifshitz}\ and\ \citenamefont
  {Pitaevskii}(1980)}]{lifshitz_statistical_1980}%
  \BibitemOpen
  \bibfield  {author} {\bibinfo {author} {\bibfnamefont {E.~M.}\ \bibnamefont
  {Lifshitz}}\ and\ \bibinfo {author} {\bibfnamefont {L.~P.}\ \bibnamefont
  {Pitaevskii}},\ }\href@noop {} {\emph {\bibinfo {title} {Statistical
  {Physics}, {Part} 2}}}\ (\bibinfo  {publisher} {Butterworth-Heinemann},\
  \bibinfo {address} {Oxford},\ \bibinfo {year} {1980})\BibitemShut {NoStop}%
\bibitem [{\citenamefont {Haldane}(1981)}]{haldane_luttinger_1981}%
  \BibitemOpen
  \bibfield  {author} {\bibinfo {author} {\bibfnamefont {F.~D.~M.}\
  \bibnamefont {Haldane}},\ }\bibfield  {title} {\bibinfo {title} {'{Luttinger}
  liquid theory' of one-dimensional quantum fluids. {I}. {Properties} of the
  {Luttinger} model and their extension to the general {1D} interacting
  spinless {Fermi} gas},\ }\href {https://doi.org/10.1088/0022-3719/14/19/010}
  {\bibfield  {journal} {\bibinfo  {journal} {J. Phys. C: Solid State Phys.}\
  }\textbf {\bibinfo {volume} {14}},\ \bibinfo {pages} {2585} (\bibinfo {year}
  {1981})}\BibitemShut {NoStop}%
\bibitem [{\citenamefont {Giamarchi}(2004)}]{giamarchi_quantum_2004}%
  \BibitemOpen
  \bibfield  {author} {\bibinfo {author} {\bibfnamefont {T.}~\bibnamefont
  {Giamarchi}},\ }\href@noop {} {\emph {\bibinfo {title} {Quantum physics in
  one dimension}}}\ (\bibinfo  {publisher} {Clarendon},\ \bibinfo {address}
  {Oxford},\ \bibinfo {year} {2004})\BibitemShut {NoStop}%
\bibitem [{\citenamefont {Kane}\ and\ \citenamefont
  {Fisher}(1992)}]{kane_transmission_1992}%
  \BibitemOpen
  \bibfield  {author} {\bibinfo {author} {\bibfnamefont {C.~L.}\ \bibnamefont
  {Kane}}\ and\ \bibinfo {author} {\bibfnamefont {M.~P.~A.}\ \bibnamefont
  {Fisher}},\ }\bibfield  {title} {\bibinfo {title} {Transmission through
  barriers and resonant tunneling in an interacting one-dimensional electron
  gas},\ }\href {https://doi.org/10.1103/PhysRevB.46.15233} {\bibfield
  {journal} {\bibinfo  {journal} {Phys. Rev. B}\ }\textbf {\bibinfo {volume}
  {46}},\ \bibinfo {pages} {15233} (\bibinfo {year} {1992})}\BibitemShut
  {NoStop}%
\bibitem [{\citenamefont {Furusaki}\ and\ \citenamefont
  {Nagaosa}(1993)}]{furusaki_single-barrier_1993}%
  \BibitemOpen
  \bibfield  {author} {\bibinfo {author} {\bibfnamefont {A.}~\bibnamefont
  {Furusaki}}\ and\ \bibinfo {author} {\bibfnamefont {N.}~\bibnamefont
  {Nagaosa}},\ }\bibfield  {title} {\bibinfo {title} {Single-barrier problem
  and {Anderson} localization in a one-dimensional interacting electron
  system},\ }\href {https://doi.org/10.1103/PhysRevB.47.4631} {\bibfield
  {journal} {\bibinfo  {journal} {Phys. Rev. B}\ }\textbf {\bibinfo {volume}
  {47}},\ \bibinfo {pages} {4631} (\bibinfo {year} {1993})}\BibitemShut
  {NoStop}%
\bibitem [{\citenamefont {Luttinger}(1963)}]{luttinger_exactly_1963}%
  \BibitemOpen
  \bibfield  {author} {\bibinfo {author} {\bibfnamefont {J.~M.}\ \bibnamefont
  {Luttinger}},\ }\bibfield  {title} {\bibinfo {title} {An {Exactly} {Soluble}
  {Model} of a {Many}-{Fermion} {System}},\ }\href
  {https://doi.org/doi:10.1063/1.1704046} {\bibfield  {journal} {\bibinfo
  {journal} {J. Math. Phys.}\ }\textbf {\bibinfo {volume} {4}},\ \bibinfo
  {pages} {1154} (\bibinfo {year} {1963})}\BibitemShut {NoStop}%
\bibitem [{\citenamefont {Rozhkov}(2005)}]{rozhkov_fermionic_2005}%
  \BibitemOpen
  \bibfield  {author} {\bibinfo {author} {\bibfnamefont {A.~V.}\ \bibnamefont
  {Rozhkov}},\ }\bibfield  {title} {\bibinfo {title} {Fermionic quasiparticle
  representation of {Tomonaga}-{Luttinger} {Hamiltonian}},\ }\href
  {https://doi.org/10.1140/epjb/e2005-00312-3} {\bibfield  {journal} {\bibinfo
  {journal} {Eur. Phys. J. B}\ }\textbf {\bibinfo {volume} {47}},\ \bibinfo
  {pages} {193} (\bibinfo {year} {2005})}\BibitemShut {NoStop}%
\bibitem [{\citenamefont {Pustilnik}\ and\ \citenamefont
  {Matveev}(2015{\natexlab{a}})}]{pustilnik_solitons_2015}%
  \BibitemOpen
  \bibfield  {author} {\bibinfo {author} {\bibfnamefont {M.}~\bibnamefont
  {Pustilnik}}\ and\ \bibinfo {author} {\bibfnamefont {K.~A.}\ \bibnamefont
  {Matveev}},\ }\bibfield  {title} {\bibinfo {title} {Solitons in a
  one-dimensional {Wigner} crystal},\ }\href
  {https://doi.org/10.1103/PhysRevB.91.165416} {\bibfield  {journal} {\bibinfo
  {journal} {Phys. Rev. B}\ }\textbf {\bibinfo {volume} {91}},\ \bibinfo
  {pages} {165416} (\bibinfo {year} {2015}{\natexlab{a}})}\BibitemShut
  {NoStop}%
\bibitem [{\citenamefont {Pustilnik}\ and\ \citenamefont
  {Matveev}(2015{\natexlab{b}})}]{pustilnik_fate_2015}%
  \BibitemOpen
  \bibfield  {author} {\bibinfo {author} {\bibfnamefont {M.}~\bibnamefont
  {Pustilnik}}\ and\ \bibinfo {author} {\bibfnamefont {K.~A.}\ \bibnamefont
  {Matveev}},\ }\bibfield  {title} {\bibinfo {title} {Fate of classical
  solitons in one-dimensional quantum systems},\ }\href
  {https://doi.org/10.1103/PhysRevB.92.195146} {\bibfield  {journal} {\bibinfo
  {journal} {Phys. Rev. B}\ }\textbf {\bibinfo {volume} {92}},\ \bibinfo
  {pages} {195146} (\bibinfo {year} {2015}{\natexlab{b}})}\BibitemShut
  {NoStop}%
\bibitem [{\citenamefont {Sasaki}\ and\ \citenamefont
  {Yamanaka}(1987)}]{sasaki_field_1987}%
  \BibitemOpen
  \bibfield  {author} {\bibinfo {author} {\bibfnamefont {R.}~\bibnamefont
  {Sasaki}}\ and\ \bibinfo {author} {\bibfnamefont {I.}~\bibnamefont
  {Yamanaka}},\ }\bibfield  {title} {\bibinfo {title} {Field theoretical
  construction of an infinite set of quantum commuting operators related with
  soliton equations},\ }\href {http://projecteuclid.org/euclid.cmp/1104116631}
  {\bibfield  {journal} {\bibinfo  {journal} {Comm. Math. Phys.}\ }\textbf
  {\bibinfo {volume} {108}},\ \bibinfo {pages} {691} (\bibinfo {year}
  {1987})}\BibitemShut {NoStop}%
\bibitem [{\citenamefont {Pogrebkov}(2003)}]{pogrebkov_boson-fermion_2003}%
  \BibitemOpen
  \bibfield  {author} {\bibinfo {author} {\bibfnamefont {A.~K.}\ \bibnamefont
  {Pogrebkov}},\ }\bibfield  {title} {\bibinfo {title} {Boson-fermion
  correspondence and quantum integrable and dispersionless models},\ }\href
  {https://doi.org/10.1070/RM2003v058n05ABEH000668} {\bibfield  {journal}
  {\bibinfo  {journal} {Russ. Math. Surv.}\ }\textbf {\bibinfo {volume} {58}},\
  \bibinfo {pages} {1003} (\bibinfo {year} {2003})}\BibitemShut {NoStop}%
\bibitem [{\citenamefont {Halperin}(1982)}]{halperin_quantized_1982}%
  \BibitemOpen
  \bibfield  {author} {\bibinfo {author} {\bibfnamefont {B.~I.}\ \bibnamefont
  {Halperin}},\ }\bibfield  {title} {\bibinfo {title} {Quantized {Hall}
  conductance, current-carrying edge states, and the existence of extended
  states in a two-dimensional disordered potential},\ }\href
  {https://doi.org/10.1103/PhysRevB.25.2185} {\bibfield  {journal} {\bibinfo
  {journal} {Phys. Rev. B}\ }\textbf {\bibinfo {volume} {25}},\ \bibinfo
  {pages} {2185} (\bibinfo {year} {1982})}\BibitemShut {NoStop}%
\bibitem [{two()}]{twofootnote}%
  \BibitemOpen
  \href@noop {} {}\bibinfo {note} {For an interaction potential that falls off
  as a power of the distance $x$ between the fermions, $U(x)\propto
  1/|x|^\lambda$ at $|x|\to\infty$, this condition requires
  $\lambda>3$.}\BibitemShut {Stop}%
\bibitem [{\citenamefont {Lieb}\ and\ \citenamefont
  {Liniger}(1963)}]{lieb_exact_1963}%
  \BibitemOpen
  \bibfield  {author} {\bibinfo {author} {\bibfnamefont {E.~H.}\ \bibnamefont
  {Lieb}}\ and\ \bibinfo {author} {\bibfnamefont {W.}~\bibnamefont {Liniger}},\
  }\bibfield  {title} {\bibinfo {title} {Exact {Analysis} of an {Interacting}
  {Bose} {Gas}. {I}. {The} {General} {Solution} and the {Ground} {State}},\
  }\href {https://doi.org/10.1103/PhysRev.130.1605} {\bibfield  {journal}
  {\bibinfo  {journal} {Phys. Rev.}\ }\textbf {\bibinfo {volume} {130}},\
  \bibinfo {pages} {1605} (\bibinfo {year} {1963})}\BibitemShut {NoStop}%
\bibitem [{\citenamefont {Sutherland}(1978)}]{sutherland_brief_1978}%
  \BibitemOpen
  \bibfield  {author} {\bibinfo {author} {\bibfnamefont {B.}~\bibnamefont
  {Sutherland}},\ }\bibfield  {title} {\bibinfo {title} {A brief history of the
  quantum soliton with new results on the quantization of the {Toda} lattice},\
  }\href {https://doi.org/10.1216/RMJ-1978-8-1-413} {\bibfield  {journal}
  {\bibinfo  {journal} {Rocky Mount. J. of Math.}\ }\textbf {\bibinfo {volume}
  {8}},\ \bibinfo {pages} {413} (\bibinfo {year} {1978})}\BibitemShut {NoStop}%
\bibitem [{\citenamefont {Sutherland}(2004)}]{sutherland_beautiful_2004}%
  \BibitemOpen
  \bibfield  {author} {\bibinfo {author} {\bibfnamefont {B.}~\bibnamefont
  {Sutherland}},\ }\href@noop {} {\emph {\bibinfo {title} {Beautiful {Models}:
  70 {Years} of {Exactly} {Solved} {Quantum} {Many}-body {Problems}}}}\
  (\bibinfo  {publisher} {World Scientific},\ \bibinfo {address} {Singapore},\
  \bibinfo {year} {2004})\BibitemShut {NoStop}%
\bibitem [{one()}]{onefootnote}%
  \BibitemOpen
  \href@noop {} {}\bibinfo {note} {For details, see Supplemental
  Material.}\BibitemShut {Stop}%
\bibitem [{\citenamefont {Lamb}(1980)}]{lamb_elements_1980}%
  \BibitemOpen
  \bibfield  {author} {\bibinfo {author} {\bibfnamefont {G.~L.}\ \bibnamefont
  {Lamb}},\ }\href@noop {} {\emph {\bibinfo {title} {Elements of soliton
  theory}}}\ (\bibinfo  {publisher} {Wiley},\ \bibinfo {address} {New York},\
  \bibinfo {year} {1980})\BibitemShut {NoStop}%
\bibitem [{\citenamefont {Imambekov}\ and\ \citenamefont
  {Glazman}(2009)}]{imambekov_phenomenology_2009}%
  \BibitemOpen
  \bibfield  {author} {\bibinfo {author} {\bibfnamefont {A.}~\bibnamefont
  {Imambekov}}\ and\ \bibinfo {author} {\bibfnamefont {L.~I.}\ \bibnamefont
  {Glazman}},\ }\bibfield  {title} {\bibinfo {title} {Phenomenology of
  {One}-{Dimensional} {Quantum} {Liquids} {Beyond} the {Low}-{Energy}
  {Limit}},\ }\href {https://doi.org/10.1103/PhysRevLett.102.126405} {\bibfield
   {journal} {\bibinfo  {journal} {Phys. Rev. Lett.}\ }\textbf {\bibinfo
  {volume} {102}},\ \bibinfo {pages} {126405} (\bibinfo {year}
  {2009})}\BibitemShut {NoStop}%
\bibitem [{\citenamefont {Imambekov}\ \emph {et~al.}(2012)\citenamefont
  {Imambekov}, \citenamefont {Schmidt},\ and\ \citenamefont
  {Glazman}}]{imambekov_one-dimensional_2012}%
  \BibitemOpen
  \bibfield  {author} {\bibinfo {author} {\bibfnamefont {A.}~\bibnamefont
  {Imambekov}}, \bibinfo {author} {\bibfnamefont {T.~L.}\ \bibnamefont
  {Schmidt}},\ and\ \bibinfo {author} {\bibfnamefont {L.~I.}\ \bibnamefont
  {Glazman}},\ }\bibfield  {title} {\bibinfo {title} {One-dimensional quantum
  liquids: {Beyond} the {Luttinger} liquid paradigm},\ }\href
  {https://doi.org/10.1103/RevModPhys.84.1253} {\bibfield  {journal} {\bibinfo
  {journal} {Rev. Mod. Phys.}\ }\textbf {\bibinfo {volume} {84}},\ \bibinfo
  {pages} {1253} (\bibinfo {year} {2012})}\BibitemShut {NoStop}%
\bibitem [{\citenamefont {Iordanskii}\ and\ \citenamefont
  {Pitaevskii}(1978)}]{iordanskii_properties_1978}%
  \BibitemOpen
  \bibfield  {author} {\bibinfo {author} {\bibfnamefont {S.~V.}\ \bibnamefont
  {Iordanskii}}\ and\ \bibinfo {author} {\bibfnamefont {L.~P.}\ \bibnamefont
  {Pitaevskii}},\ }\bibfield  {title} {\bibinfo {title} {Properties of the
  endpoint of a multiphonon spectrum},\ }\href@noop {} {\bibfield  {journal}
  {\bibinfo  {journal} {JETP Lett.}\ }\textbf {\bibinfo {volume} {27}},\
  \bibinfo {pages} {621} (\bibinfo {year} {1978})}\BibitemShut {NoStop}%
\bibitem [{\citenamefont {Matveev}(2022)}]{matveev_spectral_2022}%
  \BibitemOpen
  \bibfield  {author} {\bibinfo {author} {\bibfnamefont {K.~A.}\ \bibnamefont
  {Matveev}},\ }\bibfield  {title} {\bibinfo {title} {Spectral {Function} of
  the {Chiral} {One}-{Dimensional} {Fermi} {Liquid} in the {Regime} of {Strong}
  {Interactions}},\ }\href {https://doi.org/10.1103/PhysRevLett.128.176802}
  {\bibfield  {journal} {\bibinfo  {journal} {Phys. Rev. Lett.}\ }\textbf
  {\bibinfo {volume} {128}},\ \bibinfo {pages} {176802} (\bibinfo {year}
  {2022})}\BibitemShut {NoStop}%
\bibitem [{\citenamefont {Martin}\ and\ \citenamefont
  {Matveev}(2022)}]{martin_scar_2022}%
  \BibitemOpen
  \bibfield  {author} {\bibinfo {author} {\bibfnamefont {I.}~\bibnamefont
  {Martin}}\ and\ \bibinfo {author} {\bibfnamefont {K.~A.}\ \bibnamefont
  {Matveev}},\ }\bibfield  {title} {\bibinfo {title} {Scar states in a system
  of interacting chiral fermions},\ }\href
  {https://doi.org/10.1103/PhysRevB.105.045119} {\bibfield  {journal} {\bibinfo
   {journal} {Phys. Rev. B}\ }\textbf {\bibinfo {volume} {105}},\ \bibinfo
  {pages} {045119} (\bibinfo {year} {2022})}\BibitemShut {NoStop}%
\end{thebibliography}
\end{document}